\documentclass[twocolumn]{aastex701}
\usepackage{amsmath}
\usepackage{amssymb}
\usepackage{tipa}
\usepackage{CJKutf8}
\usepackage{wasysym}
\usepackage{combelow}
\usepackage{graphicx}
\usepackage{subcaption}
\usepackage{tcolorbox}
\tcbuselibrary{breakable}
\usepackage{booktabs}
\usepackage{multirow} 
\usepackage{array}
\usepackage{tabularx}
\usepackage{colortbl}
\usepackage{dcolumn}
\usepackage{makecell}

\shorttitle{LLM Reconstruction of Astrophysical Methods}
\shortauthors{Lin et al.}

\begin{document}

\title{Quantifying the Reproducibility of Astrophysical Methods with Large Language Models and Information Theory: A Case Study in Spectral Reconstruction}

\author[0000-0001-7737-6784]{Hsing~Wen~Lin$^{*}$ (\begin{CJK*}{UTF8}{bkai}
林省文\end{CJK*})}
\affiliation{Department of Physics, University of Michigan, Ann Arbor, MI 48109, USA}
\affiliation{Michigan Institute for Data and AI in Society, University of Michigan, Ann Arbor, MI 48109, USA}
\email{hsingwel@umich.edu}
\altaffiliation{Corresponding author}

\author[]{Zong-Fu Sie (\begin{CJK*}{UTF8}{bkai}
謝宗富\end{CJK*})}
\affiliation{Independent Researcher, Taiwan}
\email{rockzerox0010910@gmail.com}

\begin{abstract}
Modern astrophysical studies rely heavily on complex data analysis pipelines; however, published descriptions often lack the detail required for computational reproducibility. In this work, we present an information-theoretic framework to quantify how effectively a method can be reconstructed from its written description. By treating algorithmic reconstruction as a probability distribution generated by Large Language Models (LLMs), we utilize Shannon entropy and Jensen–Shannon divergence to measure how strongly text constrains the hypothesis space of valid implementations.
We demonstrate this approach through a case study of Trans-Neptunian Object (TNO) spectral reconstruction from sparse photometry. By prompting frontier LLMs with varying levels of manuscript text (Title, Abstract, and Methods), we find that while increasing text successfully clarifies the overall algorithmic structure, it fails to eliminate variance at the implementation level. This persistent variance establishes an ``entropy floor,'' demonstrating that multiple divergent implementations remain consistent with explicit instructions.
To evaluate practical reproducibility, we convert these reconstructed algorithms into executable pipelines. Our results reveal that, while LLMs easily recover core functional methodologies, they systematically fail to infer the tacit expert knowledge required for strict scientific calibration. This pilot study demonstrates that LLMs can be repurposed as a zero-shot diagnostic tool to audit methodological transparency, helping authors identify missing structural constraints and preserve scientific integrity in an era of automated research.
\end{abstract}


\section{Introduction}

Reproducibility is essential to science, yet many published results remain difficult to replicate \citep[e.g.][]{Brodeur2026, Miske2026}. Large-scale surveys point to several key drivers of irreproducibility, including selective reporting, publication pressure, weak statistical rigor, and inadequate oversight \citep{baker2016scientists}. This suggests that the problem stems from a combination of experimental design, social incentives, and methodological practices.

Specifically, vague or incomplete methodological descriptions are a solvable part of this problem. Even when results are reported accurately, missing details about computational procedures, parameters, and implementation can prevent others from replicating the work. Unlike social or cultural factors, the information within a paper can be directly analyzed and quantified.

Despite the growing complexity of data pipelines, reproducibility in astronomy and astrophysics has received relatively little systematic study. While a decade-long community effort has successfully promoted open-source practices \citep{Allen2024, Allen2018ApJS, Allen2013ASPC, Allen2013, Shamir2013, Weiner2009}, sharing code alone does not guarantee reproducibility. Software is often unavailable; even when public, it may rely on hidden assumptions or specific configurations that hinder its accessibility and reuse.

Recent advances in Large Language Models (LLMs) offer new ways to address these challenges. By generating code from natural language descriptions \citep{alphacode, codex}, LLMs provide a means to reconstruct scientific methods directly from published papers \citep{Paper2Code}. If successful, this approach would significantly increase transparency and accessibility in astronomical research. 

However, the rapid integration of generative AI into astronomical workflows has sparked intense community debate regarding the limits of automated research and the true nature of scientific understanding \citep{Peiris2026, Ting2026, NatAstronEd2026}. As recent perspectives highlight, LLMs act as a powerful diagnostic mirror for the field. While they excel at executing well-structured computational tasks, they expose the profound extent to which functional science relies on tacit, domain-specific physical intuition that is rarely formalized in print \citep{Peiris2026, Ting2026}. From this viewpoint, if a frontier LLM cannot reconstruct a pipeline from a manuscript, the failure may not lie with the model's capabilities, but rather indicates that the necessary informational constraints are fundamentally absent from the text itself. While completely disentangling LLM capabilities from textual limits remains an open challenge, the observation that multiple distinct frontier models converge on the exact same structural omissions strongly implies a fundamental absence of information in the text, rather than isolated model hallucinations.

Measuring this informational gap is central to the broader challenge of quantifying knowledge. \citet{Fanelli2019} suggests that scientific knowledge reduces uncertainty by linking data with an explanatory framework. In this view, knowledge acts as information compression: a successful theory or method narrows the range of outcomes consistent with the data. However, this model assumes the explanatory framework is already known, whereas in many cases, it must be extracted from the text itself.

In practice, the explanatory framework itself must often be inferred during text-based reconstruction. Since LLMs operate using a vast prior of potential solutions, the role of the input text is to narrow this space toward the intended approach. This suggests that the core challenge lies not only in the amount of information provided, but also in identifying the correct framework within a broad hypothesis space.

While existing frameworks quantify knowledge as information compression \citep{Fanelli2019}, we still lack a way to measure how much information a paper provides for reconstructing its method. Furthermore, there is no quantitative approach to characterize how this information constrains the range of possible solutions. This gap makes it difficult to assess a study’s reproducibility, even when all documentation is available.

In this work, we present an information-theoretic framework to quantify the information in scientific texts and the feasibility of reconstructing their methods. By treating reconstruction as a probability distribution over algorithms generated by LLMs, we measure uncertainty using Shannon entropy ($H$) and alignment using Jensen–Shannon divergence ($JSD$). We apply this framework to an astrophysical case study in spectral reconstruction, demonstrating how textual completeness limits the recovery of scientific workflows from the literature.

The rest of this paper is organized as follows. Section~\ref{sec:framework} defines the information-theoretic framework used to quantify textual information. Section~\ref{sec:exp} details our experimental design for LLM-based reconstruction, while Section~\ref{sec:case} applies this approach to an astrophysical case study. We discuss the results and their implications in Section~\ref{sec:disc} and summarize our findings in Section~\ref{sec:summary}.

\section{Information-Theoretic Framework}
\label{sec:framework}

\subsection{Problem Formulation}

We frame the reconstruction of scientific methods as an inference problem over an algorithm space. Let $X$ represent the information in a scientific text, and $Y$ denote the original method. Given $X$, a large language model (LLM) generates candidate reconstructions $\hat{Y}$, which we treat as samples from a conditional distribution:

\begin{equation}
P(Y \mid X).
\end{equation}

This distribution represents the set of plausible algorithms consistent with the text. When $X$ lacks detail, $P(Y \mid X)$ is broad, reflecting a wide range of possible interpretations. In information-theoretic terms, this corresponds to high information entropy or Shannon entropy \citep[hereafter, entropy,][]{Shannon}. As we add specific constraints to $X$, the distribution becomes more concentrated, and the entropy decreases.

\subsection{The Stochastic Sampler and Inductive Bias}

In practice, the idealized distribution $P(Y \mid X)$ is accessed through an empirical sampler (in this study, a Large Language Model). To account for the fact that any computational agent introduces its own internal constraints, we define the effective distribution modeled by the agent as:
\begin{equation}
P_{\Phi, \mathcal{T}}(Y \mid X),
\end{equation}
where $\Phi$ represents the inductive bias of the agent (including its architecture and pre-trained priors) and $\mathcal{T}$ denotes the inference configuration (comprising prompting and decoding strategies).

Under this formulation, the candidate reconstructions generated by the model, denoted as $\hat{Y}$, are treated as independent samples drawn from this effective distribution:
\begin{equation}
\hat{Y} \sim P_{\Phi, \mathcal{T}}(Y \mid X).
\end{equation}
For notational convenience, $P_{\Phi, \mathcal{T}}(Y \mid X)$ is hereafter referred to as $P(Y \mid X)$, with the agent-specific parameters implicitly assumed.

\subsection{Algorithm Distribution Estimation}
\label{sec:distribution_estimation}
To estimate the effective distribution $P(Y \mid X)$, we project the ensemble of candidate reconstructions $\{\hat{Y}_i\}$ into two complementary representational spaces:

\begin{itemize}
\item \textbf{Lexical Clustering (Keyword-based):} Captures discrete choices of method components (e.g., specific algorithm names or library functions). This represents the choice of method.
\item \textbf{Semantic Clustering (Embedding-based):} Utilizes vector embeddings to reflect the structural similarity between complete solutions. This represents how the method is implemented.
\end{itemize}
By applying clustering algorithms independently within each of these spaces (as detailed in Section~\ref{sec:metric}), we partition the ensemble into $K$ discrete algorithm classes, where each class corresponds to a cluster $\mathcal{C}_k$ (for $k = 1, \dots, K$). This discretization allows us to define the empirical frequency $q_k$ as the proportion of generated samples that fall into the $k$-th cluster:
\begin{equation}
q_k = \frac{1}{N} \sum_{i=1}^N \mathbb{I}(\hat{Y}_i \in \mathcal{C}_k),
\end{equation}
where $\mathbb{I}$ is the indicator function. In our implementation, we use these empirical frequencies as the estimators for the underlying class probabilities, such that $p_k \approx q_k$. This mapping transforms the implementation manifold into a discrete probability distribution $P = \{p_1, p_2, \dots, p_K\}$, providing the formal basis for the information-theoretic measurements in the following section.

\subsection{Uncertainty: Entropy}
\label{sec:entropy}

Using the discrete probability distribution $P$ derived in Section~\ref{sec:distribution_estimation}, we calculate the entropy to quantify the uncertainty of the algorithmic reconstruction:

\begin{equation}
H(Y \mid X) = - \sum_{k} p_k \log_2 p_k,
\end{equation}
where $p_k$ is the probability of the $k$-th algorithm cluster. High entropy indicates a broad hypothesis space with many plausible methodological variations, while low entropy suggests the method is strongly constrained by the input text. We calculate two specific measures: keyword/lexical-based entropy ($H_{\text{lex}}$) and embedding/semantic-based entropy ($H_{\text{sem}}$).

Our approach builds on the semantic entropy framework \citep{farquhar2024}, which addresses the challenge of semantic equivalence. In LLMs, standard entropy metrics (specifically token-level sequence entropy) can be misleading because the model can express a single concept through diverse word sequences that differ not only in their tokens (lexical) but also in their positions within a high-dimensional embedding space.

For example, the statements ``Claude Shannon defined entropy in the context of information theory'' and ``Shannon is the father of information entropy'' are semantically identical. However, they consist of different tokens and would occupy different coordinates in an embedding manifold. A naive entropy calculation would treat these as distinct outcomes, overestimating the true uncertainty. \citet{farquhar2024} resolved this by clustering responses into equivalence classes based on their shared meaning.

We extend this logic to scientific method reconstruction, where the underlying ``meaning'' corresponds to the specific algorithmic realization. The reason is that diverse variable names, library functions, or coding structures can implement the same mathematical procedure, creating a form of algorithmic degeneracy. Our framework aims to mitigate this by clustering implementations into approximate functional equivalence classes (see Section~\ref{sec:metric}). This approach is designed to better reflect the core ambiguity of the underlying methodology by reducing the stylistic noise inherent in diverse coding implementations.

The divergence between $H_{\text{lex}}$ and $H_{\text{sem}}$ thus reflects the multi-scale nature of algorithmic uncertainty. While lexical entropy tracks the discrete selection of specific methodological identifiers, semantic entropy captures the broader variance in their structural and stylistic realization within the implementation space. Together, these metrics allow us to analyze the information flow across different levels of abstraction: from the compositional selection of discrete methodological components ($H_{\text{lex}}$) to the structural configuration of the implementation ($H_{\text{sem}}$).

\subsection{Input Hierarchy}
\label{sec:hierarchy}

To evaluate how textual constraints influence the reconstruction process, we define a sequence of inputs with increasing information content:

\begin{equation}
X_T \subset X_{TA} \subset X_{TAM},
\end{equation}
corresponding to Title-only (T), Title + Abstract (TA), and Full Methodological Text including:

\begin{itemize}
    \item \textbf{T (Title Only) --- The Problem Space:} The baseline condition ($X_T$) establishes the astrophysical challenge without providing any methodological constraints. This captures the maximum prior uncertainty of how to solve the problem.
    \item \textbf{TA (Title + Abstract) --- The Core Idea:} The intermediate condition ($X_{TA}$) introduces the high-level summary and the authors' conceptual approach, narrowing the algorithmic hypothesis space to a specific family of solutions.
    \item \textbf{TAM (Title + Abstract + Methods) --- The Attempted Solution:} The maximum information condition ($X_{TAM}$) with the full, detailed methodological description, representing the authors' complete intended computational pipeline.
\end{itemize}

\subsection{Information Gain: Mutual Information}

To quantify the empirical measure of the algorithmic information available for reconstruction by each textual stage, we utilize the concept of Information Gain, formalized as \textbf{Mutual Information, MI} $I(X; Y)$ between the scientific text $X$ and the algorithm space $Y$. Building upon the definition of Shannon entropy in Section~\ref{sec:entropy}, mutual information represents the reduction in uncertainty of the methodological implementation after observing the text of the manuscript.

\begin{equation}
I(X; Y) = H(Y) - H(Y|X),
\end{equation}
where $H(Y)$ denotes the prior uncertainty of the algorithm space and $H(Y|X)$ is the conditional entropy given the textual information. Given our nested information hierarchy ($X_T \subset X_{TA} \subset X_{TAM}$), we decompose the information flow into discrete inter-state gains. The incremental information provided by the abstract relative to the title is:

\begin{equation}
I(X_{TA}; Y | X_T) = H(Y|X_T) - H(Y|X_{TA}).
\end{equation}
Similarly, the marginal information gain from reading the full methodology after the abstract is defined as:
\begin{equation}
I(X_{TAM}; Y | X_{TA}) = H(Y|X_{TA}) - H(Y|X_{TAM}).
\end{equation}
Because the Title-only state ($X_T$) merely defines the physical problem without providing methodological constraints, $H(Y|X_T)$serves as a practical baseline for high-uncertainty conditions, rather than a true estimate of the unconstrained entropy $H(Y)$. Consequently, the total measurable information successfully transmitted by the manuscript can be captured by the total information gain:
\begin{equation}
I(X_{TAM}; Y | X_T) = H(Y|X_T) - H(Y|X_{TAM}).
\end{equation}
This aggregate term, $I(X_{TAM}; Y | X_T)$, represents the closest practical approximation to the ideal mutual information $I(X; Y) = H(Y) - H(Y|X)$. It quantifies the total volume of algorithmic ambiguity that is successfully collapsed by the authors' complete textual documentation. While Mutual Information characterizes the \textit{magnitude} of uncertainty reduction, it does not describe the \textit{direction} of the distribution's reorganization within the manifold, which is subsequently addressed using Jensen-Shannon Divergence in Section~\ref{sec:jsd}.

\subsection{Alignment: Jensen--Shannon Divergence and Cosine Distance}
\label{sec:jsd}

While entropy ($H$) measures the ``spread'' or precision of the inferred distribution, it does not indicate whether the distribution is shifting toward the correct solution or how the model's ``beliefs'' evolve as new information is introduced. Even if the total uncertainty remains stable, the content of the distribution, i.e., which algorithm clusters are prioritized, may change significantly between stages. 

To quantify these shifts, we require a distance metric between probability distributions. However, a fundamental challenge in comparing LLM-generated ensembles is ensuring that the discrete distributions are defined over a consistent support. If each information stage were clustered independently, the resulting algorithm classes would lack a common basis for comparison, making any divergence metric mathematically ill-posed.

To resolve this, we perform clustering on the joint ensemble of all candidate reconstructions across all information states:
\begin{equation}
\mathcal{E}{_\text{total}} = \mathcal{E}_T \cup \mathcal{E}_{TA} \cup \mathcal{E}_{TAM}.
\end{equation}
This procedure establishes a shared partition of the algorithmic space $\mathcal{C} = \{ \mathcal{C}_1, \dots, \mathcal{C}_K \}$. Consequently, the distributions $P_T, P_{TA},$ and $P_{TAM}$ are defined as different probability massings over the same set of discrete states, ensuring that the comparison is mathematically well-defined. Viewing the LLM as a Bayesian inference engine, highly innovative methods will cause the posterior distribution to shift significantly as new textual constraints (T, TA, TAM) are introduced, often leaving early generic guesses behind. Because these empirical ensembles may be highly non-overlapping, taking their union ensures a shared discrete support, which is mathematically necessary for the divergence metrics discussed below.

A standard choice for measuring the distance between such distributions is the Kullback–Leibler (KL) divergence:
\begin{equation}
D_{\mathrm{KL}}(P || Q) = \sum P(i) \log(P(i)/Q(i)).
\end{equation}
However, KL divergence is poorly suited for method reconstruction because it is undefined for non-overlapping supports. If a limited-information stage ($X_T$) generates algorithm clusters that are entirely absent in a later stage ($X_{TAM}$), $Q(i) = 0$ while $P(i) > 0$. This causes the divergence to blow up to infinity. Furthermore, KL divergence is asymmetric, making it less stable for comparing different research stages.

We instead employ the Jensen–Shannon Divergence (JSD):
\begin{equation}
\mathrm{JSD}(P || Q) = \frac{1}{2} D_{\mathrm{KL}}(P || M) + \frac{1}{2} D_{\mathrm{KL}}(Q || M),
\end{equation}
where $M = \frac{1}{2}(P + Q)$. Because JSD evaluates both $P$ and $Q$ against a shared mixture distribution $M$, the denominator is never zero. This ensures that the JSD is always well-defined, symmetric, and bounded, smoothly handling cases where the underlying algorithm distributions are completely non-overlapping. We apply JSD in both \textbf{Inter-stage alignment:} (shifts between information states) and Ground-truth alignment (convergence toward the target method).

The calculation of JSD requires a robustly estimated probability mass function derived from an ensemble of samples. However, in cases where a discrete distribution cannot be reliably constructed, such as when dealing with singular model realizations or when the sample size is insufficient for stable clustering, the JSD becomes ill-posed.

To maintain the ability to quantify alignment in these instances, we employ the Cosine Distance as an alternative measure within the continuous semantic manifold. For a candidate implementation vector $\mathbf{v}_P$ and a reference vector $\mathbf{v}_Q$, the distance is defined as:

\begin{equation}
d_{\text{cos}}(\mathbf{v}_P, \mathbf{v}_Q) = 1 - \frac{\mathbf{v}_P \cdot \mathbf{v}_Q}{|\mathbf{v}_P| |\mathbf{v}_Q|}.
\end{equation}
By measuring the angular separation between the model’s mean trajectory and a singular target vector, the Cosine Distance provides a cluster-independent proxy for directional convergence, ensuring robust tracking of the reconstruction's accuracy even when the reference state lacks probabilistic spread.

In this work, MI measures the compression of algorithmic uncertainty, while JSD or Cosine Distance tracks the displacement of probability mass within the hypothesis space.

\subsection{Ground-Truth Distribution}
\label{sec:GT}

We define the ground-truth method ($Y_{\mathrm{GT}}$) as the specific algorithmic procedure used by the original authors. Under this definition, the ground truth refers to the actual implementation underlying the study, rather than the textual description or the broader class of methods capable of yielding similar results.

This distinction is critical: in many cases, multiple algorithms may produce indistinguishable results. Therefore, successful reproduction (matching the output) does not necessarily imply methodological identity (matching the process). We thus distinguish between the ground-truth method and the set of functionally equivalent implementations.

For our framework, we represent the ground-truth distribution as a delta distribution:

\begin{equation}
P_{\mathrm{GT}}(Y) = \delta(Y - Y_{\mathrm{GT}}),
\end{equation}
reflecting the assumption that a single, unique method generated the reported results.

In practice, $Y_{\mathrm{GT}}$ is often not directly observable and must be approximated. We reconstruct this reference point using all available resources, including published descriptions, source code, and, where possible, direct validation with the original authors. In this context, the ground truth serves as an operational approximation of the original implementation against which all LLM reconstructions are measured.

\subsection{Residual Information and Tacit Knowledge}
\label{sec:tacit}

Even when the full textual description $X_{\text{TAM}}$ is provided, the inferred distribution $P(Y \mid X_{\text{TAM}})$ may not converge to the ground-truth method $Y_{\mathrm{GT}}$. We quantify this remaining discrepancy as the \textbf{residual divergence}:

\begin{equation}
K_{\mathrm{res}} = \mathrm{JSD}(P(Y \mid X_{\text{TAM}}) \parallel P_{\mathrm{GT}}).
\end{equation}
Within our framework, $K_{\mathrm{res}}$ measures the additional information required to recover the original implementation beyond what is explicitly encoded in the text. Importantly, a non-zero $K_{\mathrm{res}}$ does not necessarily imply a failure of reproducibility. As multiple algorithms may yield functionally equivalent results, successful reconstruction can occur even when the inferred distribution does not concentrate on $Y_{\mathrm{GT}}$. Instead, $K_{\mathrm{res}}$ reflects the degree to which the original methodological choices are \textit{under-determined} by the available description. Thus, $K_{\mathrm{res}}$ serves as a measure of method identifiability rather than mere reproducibility.

This residual information typically arises from several sources, including:
\begin{itemize}
    \item Hidden implementation details or parameters;
    \item Implicit assumptions and environment-specific configurations;
    \item Domain expertise and iterative ``trial-and-error'' procedures that are rarely fully articulated in the text.
\end{itemize}

This notion is closely related to Michael Polanyi's concept of \textit{tacit knowledge}, which refers to knowledge that is difficult to formalize or communicate explicitly \citep{polanyi1966tacit, polanyi1958}. However, within our framework, $K_{\mathrm{res}}$ is defined operationally: it quantifies the gap between the documented method and the original practice, without requiring a formal distinction between inherently inexpressible knowledge and information that is simply omitted. As a result, $K_{\mathrm{res}}$ provides a measurable link between scientific documentation and the underlying research practice, capturing the extent to which methodological knowledge remains implicit even in supposedly ``fully described'' studies.

\section{Experimental Design}
\label{sec:exp}

We evaluate the framework using three information states ($X_T, X_{TA}, X_{TAM}$) across two complementary experiments. The first quantifies information-theoretic uncertainty and alignment (Section~\ref{sec:framework}); the second evaluates functional reproducibility by executing reconstructed code. Varying inputs from title to full methods allows us to link information flow with reconstruction success. 

\subsection{Experiment 1: Distributional Analysis with Local LLMs}
\label{sec:exp1}

\subsubsection{Algorithm Generation}
\label{sec:prompt}

To approximate $P(Y \mid X)$, we generate $N=200$ samples per state using stochastic decoding on locally hosted LLMs. This ensemble captures the model’s probability mass and supports robust information-theoretic analysis.

To ensure reproducibility and avoid API variability, we use two open-weight models: \texttt{DeepSeek-R1-Distill-Qwen-14B} \citep[\texttt{DeepSeek},][]{deepseek2025r1} and \texttt{GPT-oss-20b} \citep[\texttt{GPT-oss},][]{openai2025gptoss}. These provide complementary architectures: \texttt{DeepSeek} (reasoning-oriented) and \texttt{GPT-oss} (dense baseline). Operating at 14B–20B parameters demonstrates feasibility on standard hardware while ensuring observed uncertainty reflects textual limitations rather than model-specific artifacts.

We sample $P(Y \mid X)$ using temperature $T = 0.75$, inducing diversity via a \textbf{Boltzmann distribution} over logits. This preserves variance needed to estimate entropy and information flow. Lower temperatures would reduce variability and obscure uncertainty dynamics. The resulting diversity enables clustering and measurement of entropy ($H$) across information states.

\begin{figure*}[htbp]
\centering
\includegraphics[width=0.95\textwidth]{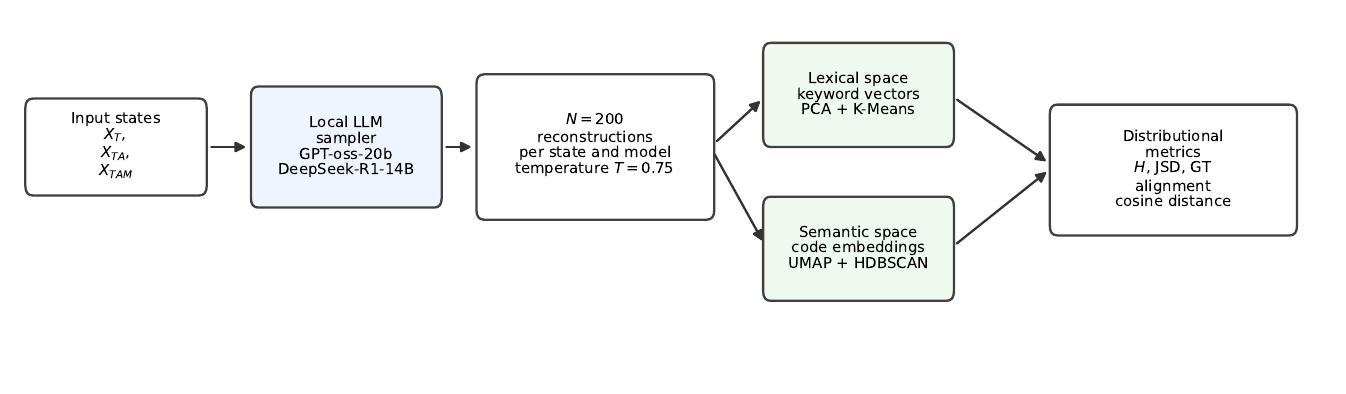}
\caption{Layout of Experiment 1. Local open-weight LLMs are used as stochastic samplers of plausible algorithmic reconstructions under three text-information states. The resulting ensembles are projected into lexical and semantic spaces before entropy, divergence, and ground-truth alignment are measured.}
\label{fig:exp1_flowchart}
\end{figure*}

LLMs are treated as zero-shot inference engines with a constrained system prompt to produce clean, mathematically representative pipelines. The full workflow is summarized in Figure~\ref{fig:exp1_flowchart}, and the key prompt controls are listed in Table~\ref{tab:exp1_prompt_controls}. The full prompt is provided in Box~\ref{box:prompt}, with \texttt{\{paper\_text\}} populated by each information state.

\begin{table*}[htbp]
\centering
\small
\caption{Prompt controls used in Experiment 1.}
\label{tab:exp1_prompt_controls}
\begin{tabular}{@{}ll@{}}
\toprule
\textbf{Control} & \textbf{Purpose} \\ \midrule
Domain anchoring & Interpret terms as computational astrophysics \\
Noise reduction & Exclude validation and plotting steps \\
Format enforcement & Return algorithm only, without reasoning text \\
Information isolation & Use zero-shot prompts with no examples \\ \bottomrule
\end{tabular}
\end{table*}

\begin{tcolorbox}[colback=gray!5!white, colframe=gray!75!black, title=System Prompt for Algorithm Extraction, label=box:prompt, arc=3mm]
\small
You are analyzing a computational astrophysics method.

Based on the following Paper description, extract a plausible computational algorithm that could implement the core reconstruction framework. Provide the algorithm in pseudocode with explicit computational steps.

Focus only on the core computational method.

\textbf{Important simplifications:}
\begin{itemize}
    \item Ignore validation procedures (LOOCV, cross validation, bootstrapping)
    \item Ignore plotting or visualization
\end{itemize}

\textbf{Important rules:}
\begin{itemize}
    \item Do NOT show reasoning or thinking.
    \item Do NOT explain the answer.
    \item Do NOT write code.
    \item Only return the final structured algorithm description.
    \item Only describe the algorithm steps.
    \item Only return the final requested output.
\end{itemize}
    
\textbf{Paper:} \\
\texttt{\{paper\_text\}}
\end{tcolorbox}

While the prompt requests pseudocode, we observed frequent departures from the specified format. \texttt{GPT-oss} consistently utilized strict pseudocode syntax, whereas \texttt{DeepSeek} often opted for detailed natural-language descriptions. However, preliminary testing revealed that the instruction to ``write pseudocode'' acted as a critical cognitive forcing function for \texttt{DeepSeek}, compelling it to provide significantly greater detail than generic extraction prompts. We kept this constraint to ensure logical depth; any syntactic variations were later resolved by semantic clustering, which prioritizes intent over format.

\subsubsection{Empirical Distribution Construction}
\label{sec:metric}

To sample the theoretical framework establish in Section~\ref{sec:framework}, we implement the two-space representation for $P(Y \mid X)$ using specific pipelines tailored to each data manifold:

\begin{itemize}
\item \textbf{Lexical (Keyword) Space:} We map each algorithm to a sparse feature vector based on a whitelist of methodological terms (the full list is provided in Appendix~\ref{app:keywords}). Given the discrete and relatively low-dimensional nature of this keyword space, we apply Principal Component Analysis (PCA) to capture the linear variance. We then utilize K-Means clustering ($K=15$) to establish a stable, fixed partition for calculating the probability mass over discrete methodological categories.
\item \textbf{Semantic (Embedding) Space:} We encode algorithms into 768-dimensional vectors using the \texttt{jina-embeddings-v2-base-code} model \citep[\texttt{Jina},][]{gunther2023jinaembeddings2}. To handle the complexity of this high-dimensional manifold, we employ Uniform Manifold Approximation and Projection \citep[\texttt{UMAP},][]{UMAP} for non-linear dimensionality reduction, preserving local topological relations. We then apply \texttt{HDBSCAN} (Hierarchical Density-Based Spatial Clustering) to identify stable consensus groups. Unlike K-Means, HDBSCAN allows for clusters of arbitrary shapes and effectively isolates outliers (noise), which is critical for filtering out non-functional or hallucinatory realizations while identifying the core semantic intent.
\end{itemize}

\subsubsection{Evaluation Metrics and Ground-Truth Alignment}
\label{sec:eval_protocol}

With the empirical distributions established, we quantify the reconstruction performance across the information states by calculating the following metrics for both the Lexical ($lex$) and Semantic ($sem$) representations:

\begin{itemize}
 \item \textbf{Algorithmic Uncertainty ($H_{lex}, H_{sem}$):} We calculate the Shannon entropy for each state to measure the convergence of the ensemble.
 \begin{itemize}
    \item \textbf{$H_{lex}$} quantifies the diversity of methodological choices (e.g., the variety of statistical estimators used).
    \item \textbf{$H_{sem}$} measures the dispersion of implementation logic within the embedding manifold.
 \end{itemize}
 \item \textbf{Internal Alignment ($\text{JSD}_{lex}, \text{JSD}_{sem}$):} We compute the inter-stage JSD to track the distribution shifts as constraints are added ($X_T \to X_{TA} \to X_{TAM}$). This quantifies the effective information flow and how each research stage prunes the hypothesis space in both keyword and semantic contexts.
 \item \textbf{Structural Alignment:} To evaluate how closely the LLM-generated algorithms recover the original astrophysics method, we measure the structural alignment using metrics tailored to the geometry of each space:
 \begin{itemize}
    \item \textbf{Lexical Alignment ($\text{JSD}_{GT}$):} In the lexical manifold, the ground-truth implementation is treated as a discrete one-hot distribution (i.e., all probability mass resides in its assigned K-Means cluster). We calculate the JSD between the generated ensemble distribution and this ground-truth distribution. This quantifies the degree of ``methodological recovery'' within the discrete space of terminology and library dependencies.
    \item \textbf{Semantic Alignment (Cosine Distance):} Because the semantic manifold is a continuous vector space where the GT exists as a single point without a defined density, we quantify alignment using geometric proximity. We calculate average cosine distance from the ensemble of generated semantic vectors to the GT vector. This measures the ``conceptual fidelity'' of the implementation logic, capturing how closely the LLM's latent reasoning matches the original scientific intent.
 \end{itemize}
\end{itemize}

The interpretation of entropy is space-dependent. In the semantic manifold, where absolute values are sensitive to clustering hyperparameters (e.g., HDBSCAN density thresholds), we prioritize relative shifts between information states as the primary signal of structural convergence. Conversely, the lexical space provides a more stable absolute measure, where the entropy magnitude reflects the inherent ambiguity of the technical vocabulary. This dual-track interpretation ensures that our analysis captures both the broad convergence trends and the specific methodological consensus.

\subsubsection{Ground-Truth Construction and Validation}
\label{sec:ground-truth-construction}

To realize the operational approximation of $P_{\mathrm{GT}}(Y)$ defined in Section~\ref{sec:GT}, we constructed a definitive reference implementation. We emphasize that, following our theoretical framework, this ``Ground-Truth'' (GT) does not represent a universal ``correct'' solution, but rather the specific methodological intent and implementation of the original study. To capture this baseline, we utilized a high-capacity model, \texttt{Gemini 2.5 Pro} \citep[\texttt{Gemini},][]{google2025gemini25pro}, with a long-context window to ingest the complete source code repository associated with the original research.

To ensure this reference point encapsulated the author's original workflow, the distilled algorithm underwent rigorous manual calibration in direct collaboration with the study's primary investigator. This human-verified anchor ensures that our $P_{\mathrm{GT}}(Y)$ represents the ``observed reality'' of the original work, providing a stable benchmark for evaluating how textual information in academic papers guides an LLM toward, or away from the specific methodology employed by the author. 

\subsection{Experiment 2: Functional Reconstruction via Frontier Models}
\label{sec:exp2}

\subsubsection{Rationale: From Information Theory to Functional Success}
\label{sec:Two-Stage}
While Experiment 1 maps the theoretical ambiguity (or ``algorithmic ellipsis'') within the manuscript’s text, scientific reproducibility ultimately demands functional validity. The core objective of this experiment is to determine whether a state-of-the-art generative AI can bridge these informational gaps to synthesize a working computational pipeline. 

\begin{figure*}[htbp]
\centering
\includegraphics[width=0.95\textwidth]{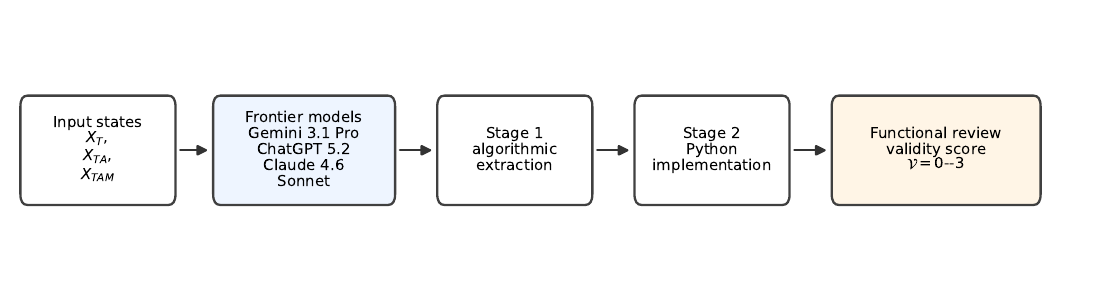}
\caption{Layout of Experiment 2. Frontier models are evaluated as autonomous reconstructors. Each information state is first converted into an algorithmic plan and then into executable Python code, which is scored using the hierarchy of scientific validity.}
\label{fig:exp2_flowchart}
\end{figure*}

To evaluate this, we transitioned from diagnostic analysis to autonomous code generation. We deployed three frontier models: \texttt{Gemini 3.1 Pro} \citep[\texttt{Gemini},][]{google2026gemini31pro}, \texttt{ChatGPT 5.2} \citep[\texttt{GPT},][]{openai2026gpt52}, and \texttt{Claude 4.6 Sonnet} \citep[\texttt{Claude},][]{anthropic2026claude46}, each tested against the same hierarchy of textual constraints ($X_T, X_{TA}, X_{TAM}$). Because executable, domain-specific Python code requires both mathematical structure and syntactic precision, we used the two-stage workflow shown in Figure~\ref{fig:exp2_flowchart}: models first extract an algorithmic plan and then translate that plan into code.

\textbf{Stage 1: Algorithmic Extraction.} The model is first required to translate the manuscript text into a structured algorithmic plan. This step anchors the model's latent space and explicitly defines the logic before any code is generated. The prompt, maintaining the noise-reduction constraints from Experiment 1 (Section~\ref{sec:prompt}), was structured as follows:

\begin{tcolorbox}[colback=gray!5!white, colframe=gray!75!black, title=System Prompt 1 --- Algorithmic Extraction, label=box:prompt1, arc=3mm]
\small
You are analyzing a computational astrophysics method. 

Based on the following paper description, extract a plausible computational algorithm that could implement the core reconstruction framework. 

\textbf{Important simplifications:}
\begin{itemize}
    \item Ignore validation procedures (LOOCV, cross validation, bootstrapping)
\end{itemize}

\textbf{Paper:} \\
\texttt{\{paper\_text\}}
\end{tcolorbox}

\textbf{Stage 2: Python Implementation.} Once the model generates the intermediate algorithmic plan\footnote{When provided with only the title ($X_T$) as context, models frequently generate a warning regarding the absence of a full description. In such instances, we bypassed this alignment safeguard by supplying a secondary reinforcement prompt (e.g., ``perform the analysis based on the title only'') to force generation based strictly on the available text.}, it is immediately passed to the second prompt. This stage instructs the model to translate its own logical framework into executable Python code. The model is tasked with constructing the full end-to-end pipeline, including data synthesis, feature extraction, and the final analytical reconstruction, ensuring that the code is modular and captures the core scientific logic described in the preceding stage.

\begin{tcolorbox}[colback=gray!5!white, colframe=gray!75!black, title=System Prompt 2 --- Code Generation, label=box:prompt2, arc=3mm]
\small
Write a Python implementation based on the algorithm. 

The code should:
\begin{enumerate}
    \item Generate synthetic spectra;
    \item Generate photometric data from synthetic spectra;
    \item Reconstruct a synthetic spectrum from photometric data.
\end{enumerate}
\end{tcolorbox}

By separating algorithmic reasoning from syntax generation, this pipeline maximizes the frontier model's ability to navigate the knowledge gap. This approach provides the most robust possible assessment of whether the published text contains sufficient information for an automated functional reproduction of the original methodology.

\subsubsection{Functional Evaluation Criteria}
\label{sec:eval}

We evaluate the functional reproducibility of the LLM-generated pipelines using the three criteria summarized in Table~\ref{tab:functional_criteria}.

\begin{table*}[htbp]
\centering
\small
\caption{Functional evaluation criteria for generated pipelines.}
\label{tab:functional_criteria}
\begin{tabular}{@{}ll@{}}
\toprule
\textbf{Criterion} & \textbf{Question} \\ \midrule
Algorithmic fidelity & Mathematical sequence and physical priors \\
Component integration & Modules and assumptions specified by the text \\
Operational validity & Executability and physically valid outputs \\ \bottomrule
\end{tabular}
\end{table*}

We intentionally avoid relying on global quantitative metrics (e.g., RMSE) between the generated and original outputs for two critical reasons. First, because the target astrophysics method is inherently probabilistic, any single execution produces a stochastic realization rather than a deterministic solution. Point-to-point numerical comparisons are therefore statistically incoherent. Second, we argue that a low numerical error can be a misleading indicator of success, potentially masking overfitting. An LLM might synthesize code that achieves a low error rate by mimicking the surface-level output of the original study, potentially through superficial pattern matching without correctly reconstructing the underlying physical mechanisms. Relying on such metrics could reward ``spurious accuracy'' while obscuring fundamental logical failures. Consequently, we prioritize qualitative algorithmic review to ensure the model has captured the ``physics of the process,'' prioritizing the integrity of the scientific reasoning over the coincidental alignment of numerical outputs.

To systematically quantify this scientific reasoning, we introduce the \textbf{Hierarchy of Scientific Validity ($V$)}, an ordinal metric ranging from 0 to 3 (Table~\ref{tab:validity_scale}). This scale assesses the degree to which an LLM reconstructs the core algorithmic framework and its essential physical constraints.

\begin{table*}[htbp]
\centering
\small
\caption{Hierarchy of Scientific Validity used to score generated code.}
\label{tab:validity_scale}
\begin{tabular}{@{}cll@{}}
\toprule
\textbf{Score} & \textbf{Label} & \textbf{Interpretation} \\ \midrule
$\mathcal{V}=0$ & Invalid / Fail & Non-executable or non-physical \\
$\mathcal{V}=1$ & Unconstrained / Partial & Executable but physically underconstrained \\
$\mathcal{V}=2$ & Functional / Overconfident & Core framework recovered; implicit priors missing \\
$\mathcal{V}=3$ & Scientific / Calibrated & Explicit and implicit constraints recovered \\ \bottomrule
\end{tabular}
\end{table*}

In our framework, a reconstruction must reach a minimum threshold of $\mathcal{V} \ge 2$ to be considered functionally LLM-reconstructable, indicating that the essential physical architecture has been successfully transmitted despite variations in specific implementation details.

\subsection{Experimental Synthesis: Bridging Information Theory and Functional Reproducibility}

The dual-track experimental architecture of this study is designed to probe two complementary facets of scientific reconstructability:

\begin{itemize}
    \item \textbf{Experiment 1 (Diagnostic Analysis):} This track quantifies the theoretical \textbf{uncertainty} and \textbf{alignment} inherent in the manuscript's text. By calculating entropy (H) and JSD across open-weight models, we mathematically measure the information density and the ``algorithmic ellipsis''--the knowledge gap--existing within the algorithmic space.
    \item \textbf{Experiment 2 (Functional Stress Test):} This track evaluates the \textbf{practical reproducibility} of the methodology. By deploying frontier generative models, we test whether state-of-the-art inference can successfully bridge the measured informational gaps to synthesize functionally valid, executable code.
\end{itemize}

Together, these experiments allow us to map abstract information-theoretic metrics directly onto observable computational outcomes. Ultimately, this framework tests a fundamental hypothesis of modern open science: whether a reduction in algorithmic entropy and a shift in semantic alignment empirically correlate with successful functional reproduction. By identifying the residual divergence ($K_{\mathrm{res}}$) between what is written and what is executed, we propose a new methodology for auditing the transparency and completeness of scientific literature.

\section{Case Study: Spectral Reconstruction from Sparse Photometry}
\label{sec:case}

To evaluate our dual-experiment framework, we select a recent astrophysical study: \textit{Probabilistic Spectral Reconstruction of Trans-Neptunian Objects from Sparse Photometry} \citep{lin2026}. This work addresses a canonical missing-information problem: reconstructing high-resolution spectra from sparse broadband photometry.

The method combines dimensionality reduction and supervised learning. First, spectra of known Trans-Neptunian Objects (TNOs) are embedded into a low-dimensional latent space via Principal Component Analysis (PCA), forming a structured prior over spectral shapes. Kernel Density Estimation (KDE) is then applied in this space to construct a continuous probability density, transforming discrete samples into a smooth manifold for interpolation. Together, PCA and KDE constrain the solution space and enable principled reconstruction from sparse inputs.

A machine learning model, specifically a Gradient Boosted Decision Tree (GBDT), is trained as an encoder mapping photometric observations to the latent PCA space. Predictions are then projected back via inverse PCA to reconstruct full spectra. This formulation separates prior structure (PCA manifold) from observational mapping (encoder), making it an ideal testbed: successful reconstruction requires identifying both components.

This case study is particularly suitable for three reasons:
\begin{enumerate}
    \item \textbf{High Algorithmic Degeneracy:} Mapping photometry to spectra is an \textit{ill-posed inverse problem} with many valid solutions (e.g., Gaussian Processes, Neural Networks, PCA-based methods). This amplifies sensitivity to missing methodological details, which manifest as high entropy in LLM reconstructions.
    
    \item \textbf{Availability of Ground Truth:} A validated reference implementation from the original authors enables direct computation of alignment, $\mathrm{JSD}(P(Y \mid X) \parallel P_{\mathrm{GT}})$, against a verified baseline.
    
    \item \textbf{Minimal Contamination Risk:} The study was \textit{unpublished} during all experiments, with no public code or preprint available. This ensures that reconstruction results reflect zero-shot inference rather than memorization. Although now published, all results were frozen prior to release.
\end{enumerate}

\subsection{Information States and Reconstruction Setup}

To prevent leakage from reported results, the $X_{TAM}$ extraction was restricted to the Methods section. Manual review confirmed that later sections contain no additional algorithmic constraints.

Following Section~\ref{sec:exp}, we applied the three information states to local LLM ensembles ($N=200$, Experiment 1) and frontier API models (Experiment 2). Reconstructed candidates were mapped to $P(Y|X)$ using the clustering pipelines in Section~\ref{sec:framework}. For functional evaluation, API-generated algorithms were implemented in Python and assessed using the criteria in Section~\ref{sec:eval}. The Ground-Truth (GT) baseline (Section~\ref{sec:metric}) is provided in Appendix~\ref{app:ground_truth}.

\subsection{Experiment 1: Algorithmic Entropy and Alignment}

Experiment 1 quantifies how methodological text constrains the LLM algorithmic hypothesis space, using Mutual Information (MI) to measure information gain and Jensen–Shannon Divergence (JSD) to assess distribution alignment. Statistical uncertainty from finite sampling of local LLM ensembles ($N=200$) is estimated via non-parametric bootstrap resampling (1,000 iterations).

Tables~\ref{tab:keyword_metrics} and \ref{tab:semantic_metrics} summarize the resulting lexical and semantic reconstruction dynamics across the three information states ($X_T$, $X_{TA}$, $X_{TAM}$), with corresponding visualizations in Figures~\ref{fig:lexical}, \ref{fig:semantic_gpt}, and \ref{fig:semantic_deepseek}.

\begin{table*}[htbp]
\centering
\caption{Experiment 1: Quantitative metrics of algorithmic reconstruction in the \textbf{Lexical Space}}
\label{tab:keyword_metrics}
\begin{tabular}{@{}llcc@{}}
\toprule
\textbf{Information-Theoretic Metric} & \textbf{State / Transition} & \textbf{GPT-OSS} & \textbf{DeepSeek} \\ \midrule
\textbf{Shannon Entropy} & \textbf{T} (Title Only) & 3.54 $\pm$ 0.06 & 3.65 $\pm$ 0.04\\
$H(Y|X)$ (bits)& \textbf{TA} (Title + Abstract) & 3.52  $\pm$ 0.04  & 3.03 $\pm$ 0.05\\
 & \textbf{TAM} (Full Method) & 3.06 $\pm$ 0.04 & 2.62 $\pm$ 0.04 \\ \midrule
\textbf{Information Gain} & $\mathbf{T \rightarrow TA}$ & 0.02 $\pm$ 0.07 & 0.62 $\pm$ 0.06 \\
$I(X_2; Y | X_1)$ (bits) & $\mathbf{TA \rightarrow TAM}$ & 0.46 $\pm$ 0.06 & 0.41 $\pm$ 0.06 \\
 & $\mathbf{T \rightarrow TAM}$ (Total) & 0.48 $\pm$ 0.07 & 1.03 $\pm$ 0.06 \\ \midrule
\textbf{Inter-State Shift} & $\mathbf{T \rightarrow TA}$ & 0.30 $\pm$ 0.02 & 0.40 $\pm$ 0.02 \\
$\mathrm{JSD}(X_1 || X_2)$  (bits)& $\mathbf{TA \rightarrow TAM}$ & 0.40 $\pm$ 0.01 & 0.22 $\pm$ 0.01 \\
 & $\mathbf{T \rightarrow TAM}$ & 0.69 $\pm$ 0.02 & 0.68 $\pm$ 0.02 \\ \midrule
\textbf{Ground-Truth Alignment} & \textbf{T} & 0.71 $\pm$ 0.03 & 0.68 $\pm$ 0.03\\
$\mathrm{JSD}(P || P_{\mathrm{GT}})$  (bits)& \textbf{TA} & 0.53 $\pm$ 0.02 & 0.36 $\pm$ 0.01 \\
 & \textbf{TAM} & 0.26 $\pm$ 0.02 & 0.16 $\pm$ 0.01\\ \bottomrule
\end{tabular}
\end{table*}

\begin{table*}[htbp]
\centering
\caption{Experiment 1: Quantitative metrics of algorithmic reconstruction in the \textbf{Semantic Space}}
\label{tab:semantic_metrics}
\begin{tabular}{@{}llcc@{}}
\toprule
\textbf{Information-Theoretic Metric} & \textbf{State / Transition} & \textbf{GPT-OSS} & \textbf{DeepSeek} \\ \midrule
\textbf{Semantic Entropy} & \textbf{T} (Title Only) & 2.0 $\pm$ 0.1 & 1.76 $\pm$ 0.08 \\
$H_{emb}(Y|X)$ (bits) & \textbf{TA} (Title + Abstract) & 2.0 $\pm$ 0.1 & 1.40 $\pm$ 0.09 \\
 & \textbf{TAM} (Full Method) & 2.0 $\pm$ 0.1 & 1.41 $\pm$ 0.08 \\ \midrule
\textbf{Information Gain} & $\mathbf{T \rightarrow TA}$ & 0.00 $\pm$ 0.14 & 0.36 $\pm$ 0.12 \\
$I(X_2; Y | X_1)$ (bits) & $\mathbf{TA \rightarrow TAM}$ & 0.00 $\pm$ 0.14 & -0.01 $\pm$ 0.12 \\
 & $\mathbf{T \rightarrow TAM}$ (Total) & 0.00 $\pm$ 0.14 & 0.35 $\pm$ 0.11 \\ \midrule
\textbf{Inter-State Shift} & $\mathbf{T \rightarrow TA}$ & 0.72 $\pm$ 0.03 & 0.87 $\pm$ 0.03 \\
$\mathrm{JSD}(X_1 || X_2)$ (bits) & $\mathbf{TA \rightarrow TAM}$ & 0.72 $\pm$ 0.03 & 0.25 $\pm$ 0.04 \\
 & $\mathbf{T \rightarrow TAM}$ & 0.79 $\pm$ 0.03 & 0.85 $\pm$ 0.03 \\ \midrule
\textbf{Ground-Truth Alignment} & \textbf{T} & 0.33 & 0.30 \\
Cosine Distance to GT & \textbf{TA} & 0.25 & 0.19 \\
 & \textbf{TAM} & 0.24 & 0.18 \\ \bottomrule
\end{tabular}
\end{table*}

\subsubsection{The Lexical Space}
\label{sec:lexical}

\begin{figure*}[t]
\centering
\includegraphics[width=2.\columnwidth]{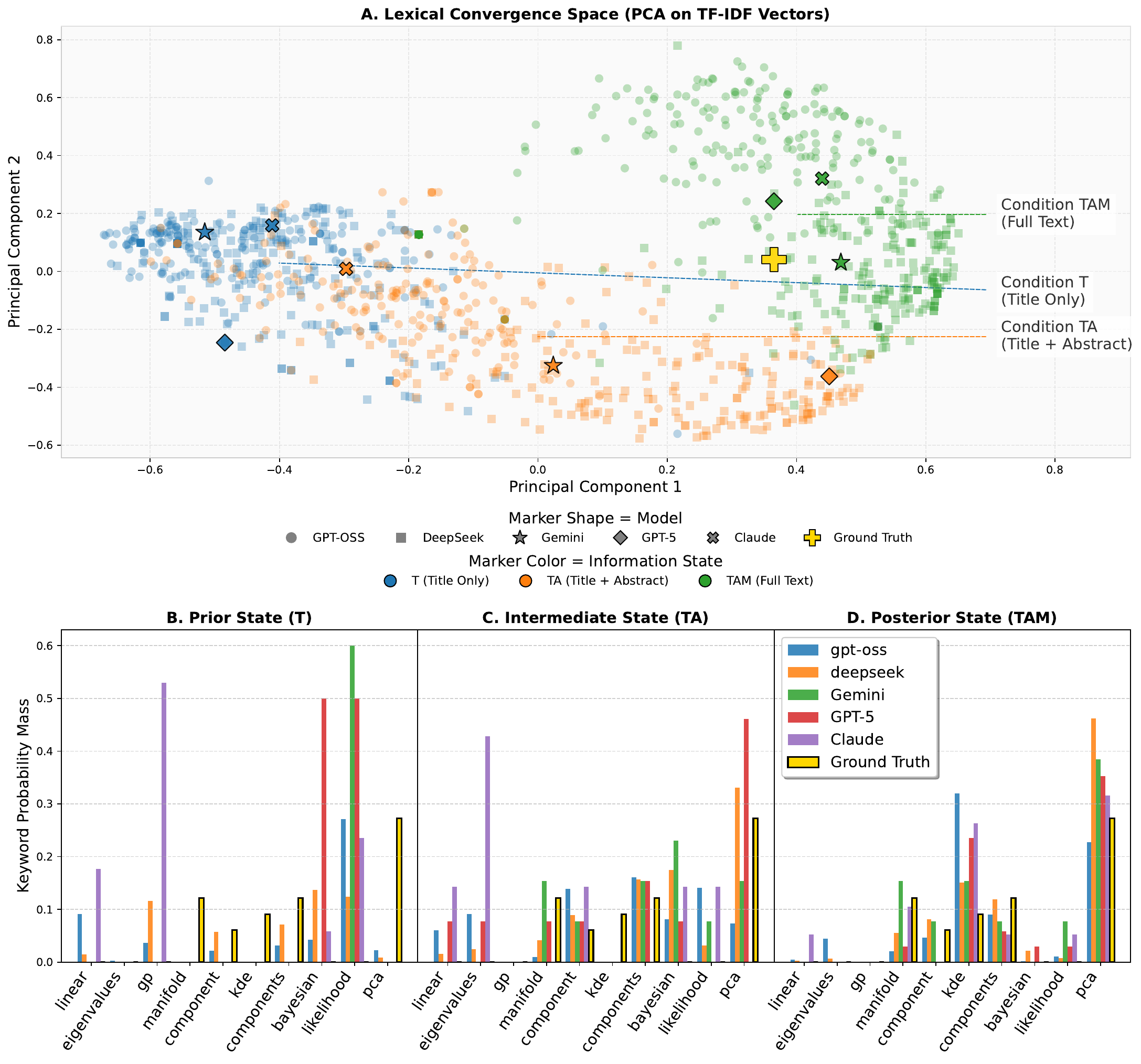}
\caption{Lexical reconstruction space across information conditions. (A) PCA projection of TF--IDF representations. Points are colored by condition (T, TA, TAM) and shaped by model. (B--D) Keyword probability distributions. The transition from T to TA produces the largest reorganization of keyword weights toward method-defining terms, while the transition from TA to TAM yields localized refinement.}
\label{fig:lexical}
\end{figure*}

Figure~\ref{fig:lexical} visualizes the lexical reconstruction landscape across information conditions using a PCA projection of TF--IDF representations. Each point denotes a reconstructed solution, categorized by information state ($X_T$, $X_{TA}$, $X_{TAM}$) and model.

At the global level (Panel A), solutions form structured clusters rather than collapsing to a point, indicating that reproducible implementations occupy a \textbf{distributed manifold} in lexical space. Even under full constraints, identical methods (e.g., PCA) admit diverse technical descriptions. Reconstructions from different LLMs occupy overlapping regions, suggesting cross-model convergence on core concepts.

The transition from Title (T) to Title+Abstract (TA) drives the primary conceptual reorientation, though with strong model dependence. As shown in Panels B and C, Title-only reconstructions rely on broad terms (e.g., \textit{Bayesian, likelihood}) and generic Gaussian frameworks. With the abstract, models resolve the problem structure, identifying eigenvalue decomposition and shifting toward PCA. This reduces divergence from ground truth; for example, \texttt{DeepSeek} decreases JSD from $0.68$ to $0.36$.

This shift is quantified by the conditional mutual information $I(X_{TA}; Y | X_T)$. \texttt{DeepSeek} gains $0.62$ bits ($3.65 - 3.03$), substantially constraining the hypothesis space, while \texttt{GPT-oss} gains only $0.02$ bits, maintaining a broader lexical distribution.

The transition from \textbf{TA to Full Methods (TAM)} (Panel C $\rightarrow$ D) primarily refines the solution. Models concentrate on ground-truth keywords, and general terminology diminishes. Techniques absent from the abstract (e.g., KDE) emerge only here. Learning dynamics diverge: \texttt{DeepSeek} gains a marginal $0.41$ bits, whereas \texttt{GPT-oss} gains $0.46$ bits, achieving its largest alignment improvement at this stage.

Overall, the full manuscript transmits $1.03$ bits of effective algorithmic information for \texttt{DeepSeek} and $0.48$ bits for \texttt{GPT-oss}. While entropy decreases with information gain, it reaches a non-zero \textbf{entropy floor} rather than collapsing. This residual dispersion reflects lexical flexibility: synonymous terminology maintains diversity even as \textbf{Ground-Truth (GT) alignment} improves (Table~\ref{tab:keyword_metrics}). Thus, MI prunes the hypothesis space without enforcing a unique representation.

Finally, information gain and distributional shift are decoupled. While MI accumulates additively, JSD evolves non-linearly, indicating that LLMs do not follow a straight path toward the ground truth. Instead, each constraint induces a global reorganization of the lexical manifold, consistent with a \textbf{curved} reconstruction landscape under an information-geometric interpretation.

\subsubsection{The Semantic Space}
\label{sec:semantic}

While the lexical analysis shows convergence toward a shared conceptual vocabulary (high mutual information in keyword adoption), the semantic space reveals how these concepts are executed. Because semantic embeddings encode logic flow and implementation structure, reconstructed solutions do not collapse to a single point. Instead, persistent dispersion reflects functionally equivalent but structurally diverse implementations.

\begin{figure*}[t]
\centering
\includegraphics[width=2\columnwidth]{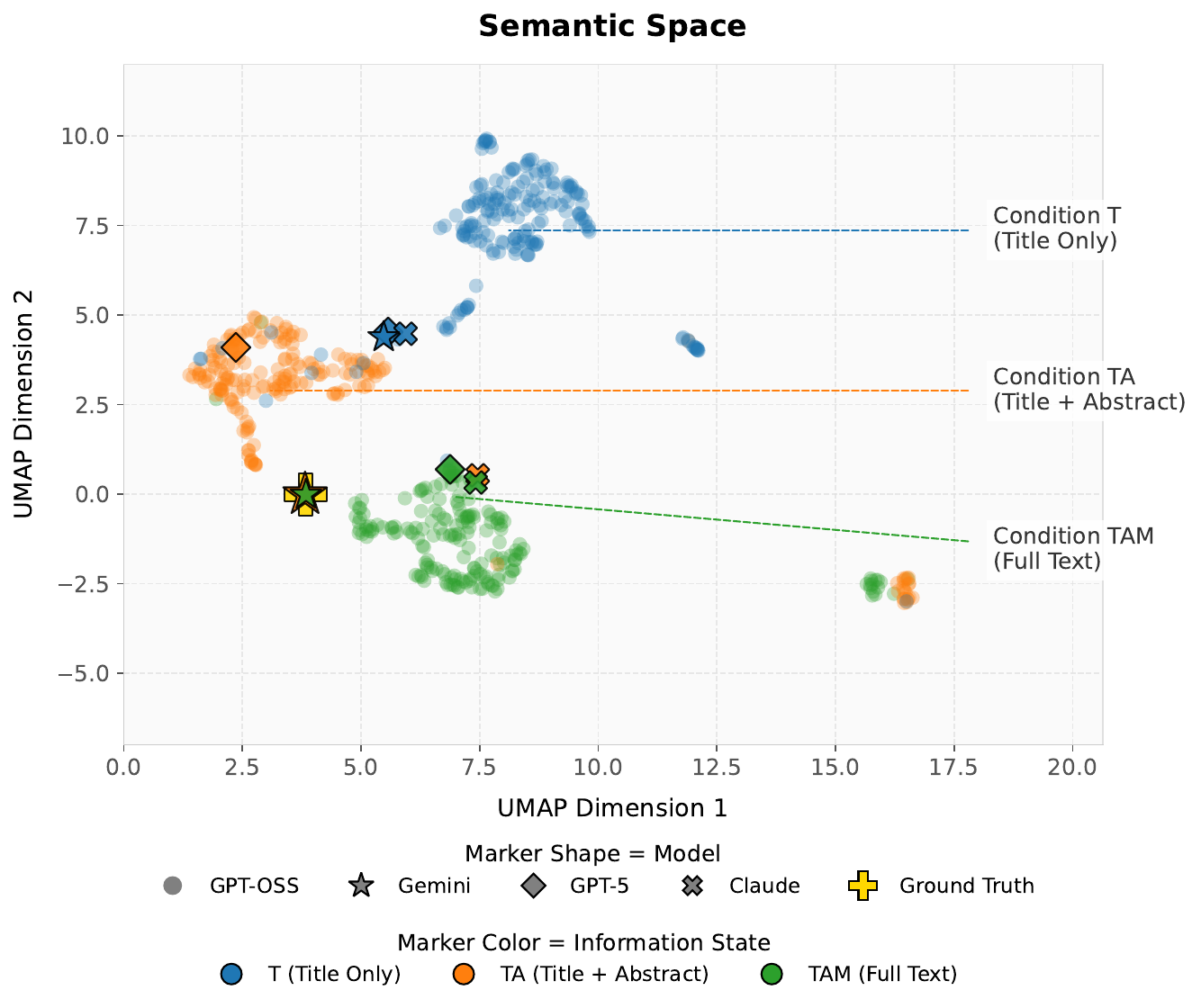}
\caption{Semantic reconstruction space for \texttt{GPT-oss}. UMAP projection of embedding representations across information states: Title (T), Title+Abstract (TA), and Full Text (TAM). The ground truth (GT) is shown for reference. The states exhibit a quasi-linear progression toward the GT, with relatively uniform spacing between information states. Note: The isolated cluster in the lower-right corner represents failed, meaningless reconstructions generated by the LLM, which are correctly isolated by the UMAP projection and should be treated as noise.}
\label{fig:semantic_gpt}
\end{figure*}

\begin{figure*}[t]
\centering
\includegraphics[width=2\columnwidth]{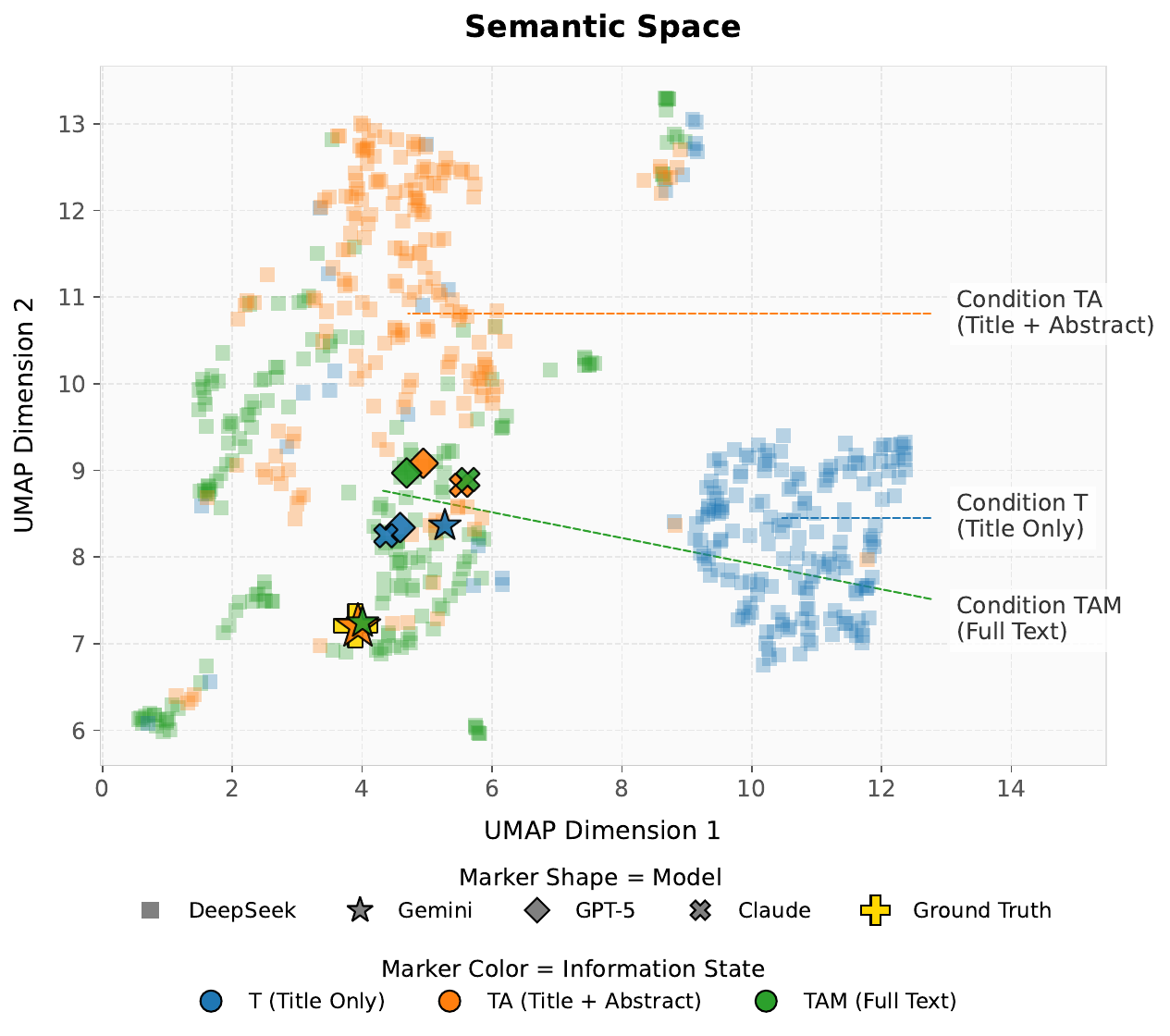}
\caption{Semantic reconstruction space for \texttt{DeepSeek}. Unlike \texttt{GPT-oss}, \texttt{DeepSeek} shows rapid semantic convergence at the TA stage, with the TAM condition providing only localized refinement. This suggests that the abstract captures the majority of the algorithmic essence for reasoning-capable models.}
\label{fig:semantic_deepseek}
\end{figure*}

Figures~\ref{fig:semantic_gpt} and \ref{fig:semantic_deepseek} show UMAP projections of semantic embeddings across information states ($X_T$, $X_{TA}$, $X_{TAM}$), with each point representing a reconstructed candidate, colored by information state and shaped by model.

At the global level, the models exhibit distinct geometric behaviors. For \texttt{GPT-oss} (Figure~\ref{fig:semantic_gpt}), the three states are well separated, showing a uniform trajectory in latent space. This is reflected in comparable inter-state JSD shifts ($\text{JSD}_{T \to TA} \approx 0.72 \pm 0.03$ bits; $\text{JSD}_{TA \to TAM} \approx 0.72 \pm 0.03$ bits). However, this movement is decoupled from information gain: semantic entropy remains fixed at $2.0$ bits, yielding $I(X_{TAM}; Y | X_T) \approx 0.0$ bits. Thus, text shifts probability mass toward the correct region (JSD) but fails to compress structural uncertainty.

In contrast, \texttt{DeepSeek} (Figure~\ref{fig:semantic_deepseek}) shows non-linear convergence with measurable information gain. The $X_T$ cluster is isolated, while $X_{TA}$ and $X_{TAM}$ overlap. The abstract induces a strong shift ($\text{JSD}_{T \to TA} \approx 0.87 \pm 0.03$ bits) and information gain of $0.36$ bits (entropy $1.76 \to 1.40$). Adding full methods produces only minor refinement ($\text{JSD}_{TA \to TAM} \approx 0.25 \pm 0.04$ bits) and no additional information gain. This indicates that, for reasoning-oriented models, the abstract contains most of the signal needed to constrain the structural hypothesis space.

Despite different trajectories, both models move toward the ground truth (GT) embedding as information increases. \texttt{DeepSeek} consistently achieves stronger alignment (lower cosine distance) than \texttt{GPT-oss} (Table~\ref{tab:semantic_metrics}).

Together, these results reveal the limit of textual transmission: a semantic \textit{entropy floor}. While lexical mutual information compresses vocabulary variance, semantic uncertainty remains largely irreducible. \texttt{GPT-oss} extracts $0.0$ bits of semantic information, and \texttt{DeepSeek} only $\sim0.35$ bits before plateauing. This contrast shows that additional text primarily \textit{reorganizes} probability mass toward the ground truth (via JSD and cosine distance), rather than eliminating structural variance (via mutual information). Even mathematically identical methods therefore retain a highly degenerate space of implementations.

Finally, as a non-linear dimensionality reduction technique, UMAP preserves local neighborhoods rather than global geometry. The 2D projections should therefore be interpreted qualitatively; quantitative comparisons rely on the information-theoretic metrics in Table~\ref{tab:semantic_metrics}.

\subsubsection{Summary of Algorithmic Reconstruction Dynamics}
\label{sec:exp1_summary}

Synthesizing results across lexical and semantic spaces, we identify four key observations governing how LLMs reconstruct scientific methodology from text:

\begin{enumerate}
    \item \textbf{Entropy Floor and Limits of Information Gain:} Algorithmic uncertainty does not collapse to zero, even under strong textual constraints. While models extract mutual information to constrain conceptual vocabulary, the semantic hypothesis space resists compression (often yielding near-zero structural information gain). Both spaces exhibit a functional ``entropy floor,'' indicating that reproducible implementations form a \textit{distributed manifold} of functionally equivalent but structurally diverse solutions, rather than a single point.

    \item \textbf{Directional Convergence vs. Compression:} Despite residual uncertainty, added information consistently drives solutions toward the Ground Truth (GT). Models reorganize probability mass toward the correct region (captured by decreasing JSD and cosine distance) even when mutual information plateaus. Text therefore primarily \textit{displaces} rather than \textit{compresses} the implementation space.
    
    \item \textbf{Primacy of the Idea:} The transition from Title to Abstract ($T \rightarrow TA$) produces the largest information gain and geometric shift, especially for reasoning-oriented models. In contrast, $TA \rightarrow TAM$ yields diminishing returns, acting as refinement rather than redirection. An exception is the lexical behavior of \texttt{GPT-oss}, where full methods induce the largest shift, highlighting architecture-dependent dynamics.

\end{enumerate}

In Experiment 2 (Section~\ref{sec:exp2}), we test whether these dynamics translate into observable software engineering behavior when frontier models generate executable Python code.

\subsection{Experiment 2: Functional Reproducibility and Code Evolution}
\label{sec:exp2_result}

While Experiment 1 characterized the landscape of algorithmic ambiguity, Experiment 2 evaluates whether this residual uncertainty prevents practical scientific reproduction. Using the functional criteria in Section~\ref{sec:eval}, we conduct a human-in-the-loop evaluation, analyzing and, where necessary, debugging LLM-generated code on real astronomical datasets.

\begin{figure*}
        \centering
        \begin{subfigure}[b]{0.49\textwidth}
            \centering
            \includegraphics[width=\textwidth]{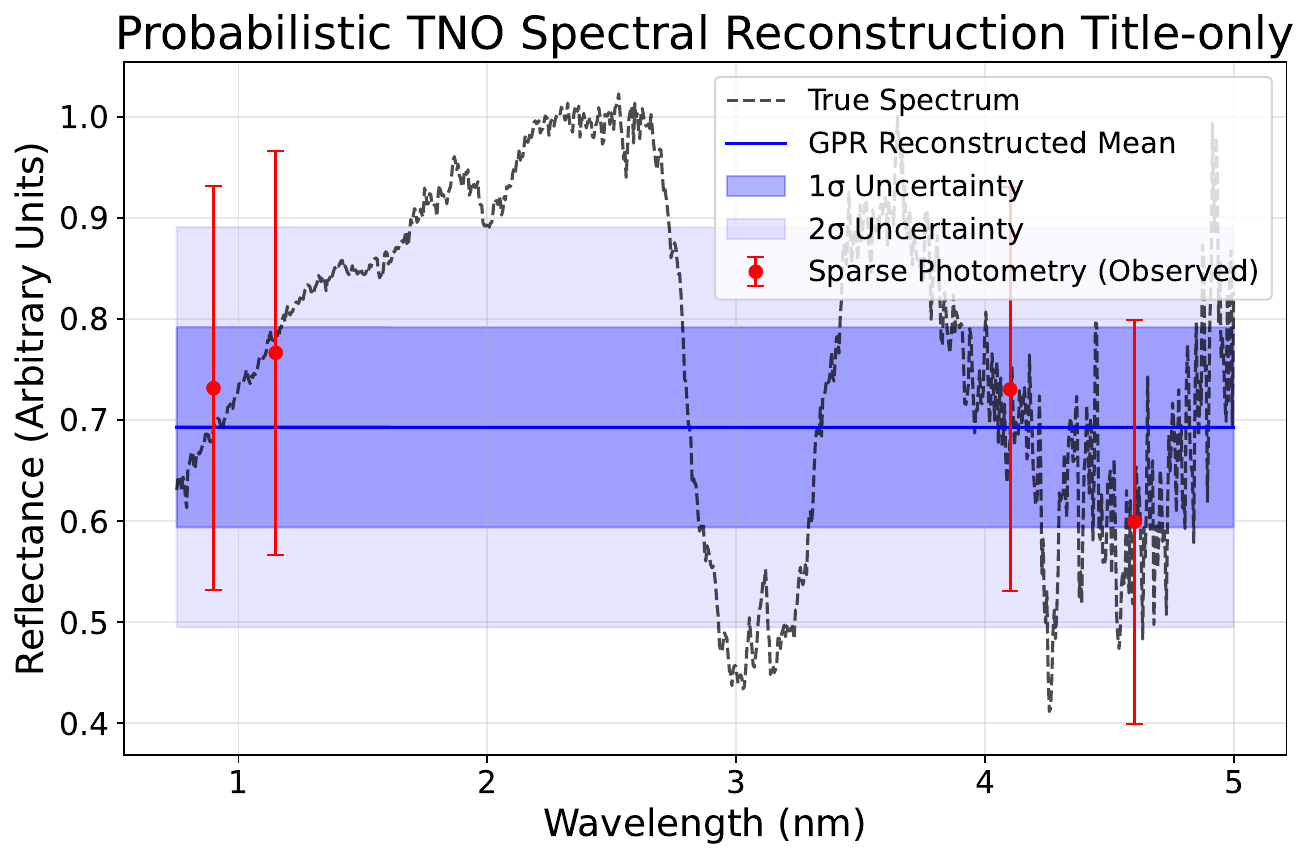}
        \end{subfigure}
        \hfill
        \begin{subfigure}[b]{0.49\textwidth}  
            \centering 
            \includegraphics[width=\textwidth]{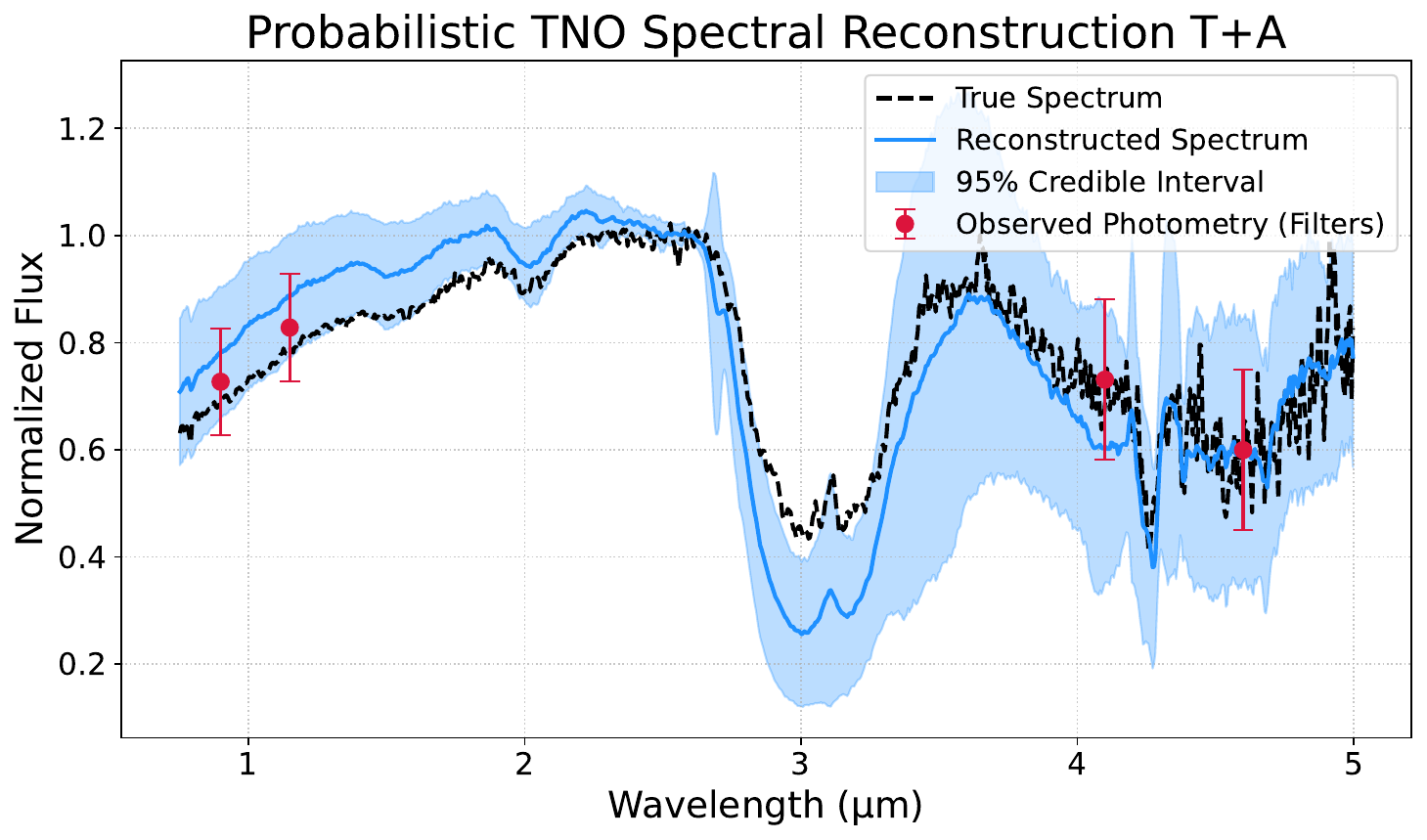}
        \end{subfigure}
        \vskip\baselineskip
        \begin{subfigure}[b]{0.49\textwidth}   
            \centering 
            \includegraphics[width=\textwidth]{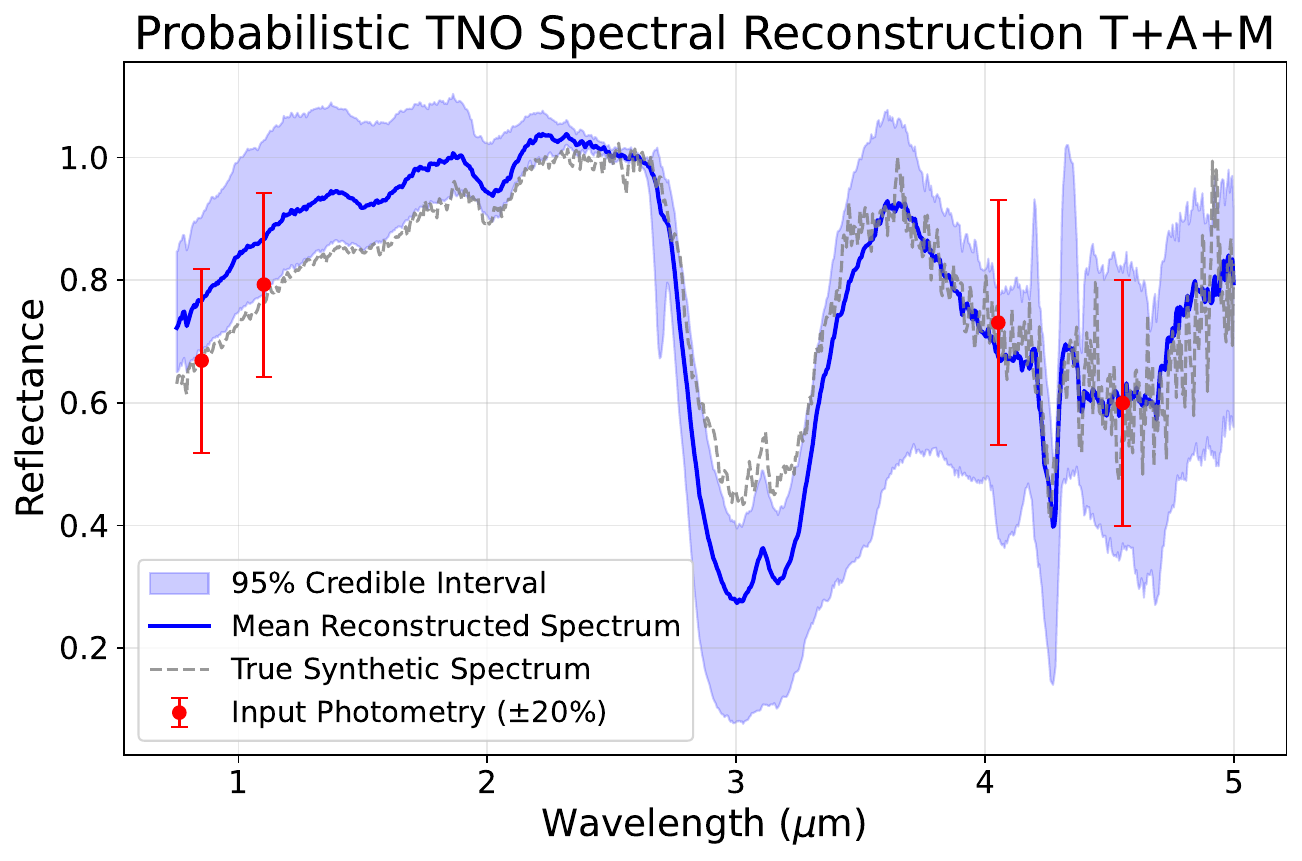}
        \end{subfigure}
        \hfill
        \begin{subfigure}[b]{0.49\textwidth}   
            \centering 
            \includegraphics[width=\textwidth]{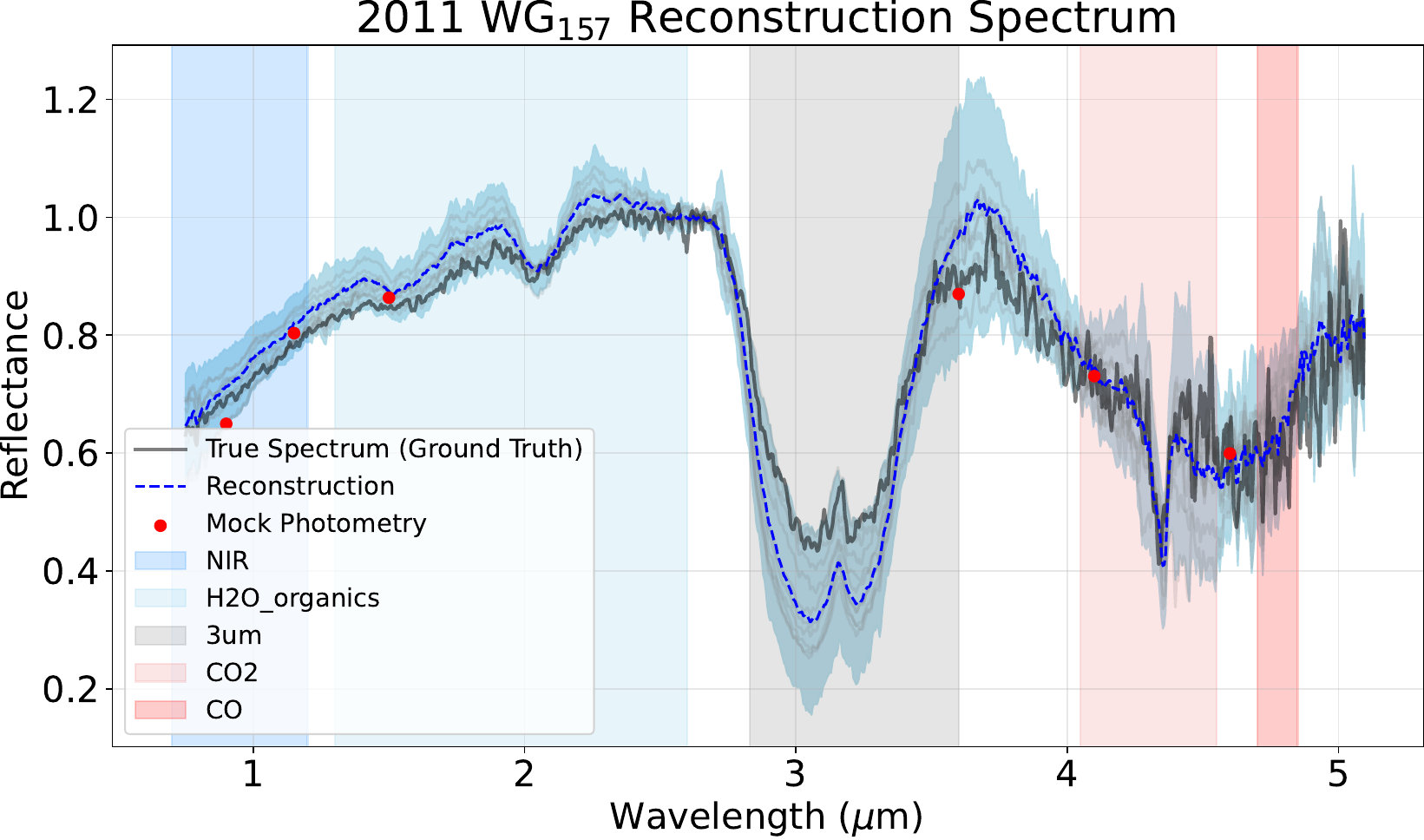}
        \end{subfigure}
        \caption{Functional spectral reconstruction of the Neptune Trojan 2010~TS$_{191}$ across the informational hierarchy. The panels demonstrate the outputs generated by LLM-synthesized pipelines under increasing textual constraints ($X_T$, $X_{TA}$, $X_{TAM}$) compared against the Ground Truth (GT). This progression visually captures the evolution of scientific validity ($\mathcal{V}$): transitioning from an invalid generic prior in $X_T$ ($\mathcal{V}=0$), to severe unconstrained overfitting in $X_{TA}$ ($\mathcal{V}=1$), and finally to a structurally viable but overconfident implementation in $X_{TAM}$ ($\mathcal{V}=2$) that approaches the calibrated GT baseline ($\mathcal{V}=3$).}
        \label{fig:spectra}
\end{figure*}

Following the two-stage prompting framework (Section~\ref{sec:Two-Stage}), frontier API models are tasked with first generating an algorithmic extraction and subsequently translating that logic into executable Python scripts. To rigorously test functional reproducibility, we employ a \textit{data substitution protocol}: replacing the synthetic spectra initially generated by the LLMs with real observational data from the original study. This procedure enables a direct comparison between the outputs of the LLM-reconstructed pipeline and the published results, ensuring that the models recover the underlying physical methodology rather than merely the software syntax.The quantitative and qualitative outcomes of these evaluations are summarized in Tables~\ref{tab:eval} and \ref{tab:reason}. Furthermore, the practical impact of this informational hierarchy is visually demonstrated in Figure~\ref{fig:spectra}, which illustrates how the stepwise addition of textual constraints ($X_T \rightarrow X_{TA} \rightarrow X_{TAM}$) progressively rescues the reconstructed spectra from severe unconstrained overfitting to functional scientific alignment.

\subsubsection{Qualitative Assessment: The Hierarchy of Scientific Validity}
\label{sec:qualitative}

At the Title-only ($X_T$) stage, all models fail to recover the correct framework, defaulting to generic priors such as Gaussian Processes (GP) or Gaussian Mixture Models (GMM). While mathematically reasonable, these lack the domain-specific logic required to constrain incomplete observations using structured physical priors. As a result, models fail to reproduce TNO spectral features and generate physically meaningless outputs (see Figure~\ref{fig:spectra}, top left panel). This reflects a mismatch of the solution idea rather than the variance of the implementation, placing all $X_T$ reconstructions in $\mathcal{V}=0$ (Invalid).

Introducing the abstract ($X_{TA}$) triggers a clear conceptual shift. \texttt{Gemini}, \texttt{GPT}, and \texttt{Claude} converge on the correct foundation: projecting photometric data into a latent space via PCA and mapping it with Bayesian Linear Regression (BLR). This produces plausible mean reconstructions by combining valid spectral basis vectors (Figure~\ref{fig:spectra}, top right panel). However, a critical flaw emerges: severe overfitting could occur due to an unconstrained latent space. Without an explicit density prior (e.g., KDE), BLR treats all PCA components as free parameters, allowing interpolation through regions of zero physical probability. The resulting spectra may violate physical constraints (e.g., negative flux). These solutions therefore fall into $\mathcal{V}=1$ (Unconstrained): structurally valid but lacking the regularization required for functional ML ($\mathcal{V}=2$).

With full methods ($X_{TAM}$), models recover the complete pipeline, converging to \textit{PCA + KDE + ML}. This transition is primarily refinement rather than redirection, consistent with the near-zero semantic information gain ($I(X_{TAM}; Y | X_{TA}) \approx 0$) observed in Experiment 1. However, convergence is not to a unique implementation. As shown in Figures~\ref{fig:lexical}, \ref{fig:semantic_gpt}, and \ref{fig:semantic_deepseek}, models remain distributed across a degenerate manifold, confirming the entropy floor.

Within this dispersion, a notable structural consensus emerges: frontier API models cluster into a shared ``API dialect,'' a common implementation style distinct from the human Ground Truth. This gap highlights the limits of textual transmission. Although models achieve functional viability ($\mathcal{V}=2$), uncertainty estimation remains systematically flawed. Most implementations ignore covariance between latent PCA components, sampling them independently and producing physically implausible spectra. In contrast, the Ground Truth achieves calibrated uncertainty ($\mathcal{V}=3$) through a principled probabilistic formulation (e.g., Gaussian Copulas).

This discrepancy exposes a limitation: while LLMs reliably recover explicit methodology, they consistently miss \textbf{implicit expert knowledge}: the domain-specific structural constraints that are rarely articulated but essential for full scientific calibration.\footnote{Notably, as a direct result of this reconstruction experiment, the authors of the subject study revised their manuscript prior to publication to explicitly include the previously missing description of the Gaussian copula.}

\begin{table*}[htbp]
\centering
\caption{Experiment 2: Algorithmic Evolution and Ground Truth Alignment}
\label{tab:eval}
\resizebox{\textwidth}{!}{%
\begin{tabular}{l l c c c c}
\toprule
\textbf{State} & \textbf{Evaluation Criteria} & \textbf{Gemini 3.1 Pro} & \textbf{GPT-5.2} & \textbf{Claude 4.6 Sonnet} & \textbf{Ground Truth} \\ \midrule

\textbf{T}           & \textbf{Core Framework}     & Gaussian Process       & Gaussian Mixture Model & Gaussian Process       & \textit{N/A} \\
\textit{(Title)}     & \textbf{Method. Components} & Reg. w/ RBF Kernel     & Expectation-Maximization & Reg. w/ Mat\'ern-3/2 & \textit{N/A} \\
                     & \textbf{Mean Reconstruction}& Fail                   & Fail                   & Fail                   & \textit{N/A} \\
                     & \textbf{Uncertainty Coverage}& \textit{Invalid}       & \textit{Invalid}       & \textit{Invalid}       & \textit{N/A} \\ \midrule

\textbf{TA}          & \textbf{Core Framework}     & PCA (No KDE Manifold)  & PCA (No KDE Manifold)  & PCA (No KDE Manifold)  & \textit{N/A} \\
\textit{(+ Abstract)}& \textbf{Method. Components} & Bayesian Linear Regression     & Bayesian Linear Regression     & Bayesian Linear Regression     & \textit{N/A} \\
                     & \textbf{Mean Reconstruction}& Partial                & Partial                   & Partial             & \textit{N/A} \\
                     & \textbf{Uncertainty Coverage}& \textit{Unconstrained} & \textit{Unconstrained} & \textit{Unconstrained} & \textit{N/A} \\ \midrule

\textbf{TAM}         & \textbf{Core Framework}     & PCA + KDE + ML         & PCA + KDE + ML         & PCA + KDE + ML         & \textbf{PCA + KDE + ML} \\
\textit{(Full Text)} & \textbf{Method. Components} & RF + Monte Carlo       & RF + Importance Samp.  & GBR + Monte Carlo      & \textbf{AutoGluon + Copula} \\
                     & \textbf{Mean Reconstruction}& \textbf{Pass}          & \textbf{Pass}          & \textbf{Pass}          & \textbf{Benchmark} \\
                     & \textbf{Uncertainty Coverage}& \textit{Overconfident} & \textit{Overconfident} & \textit{Overconfident} & \textbf{\textit{Calibrated}} \\ \bottomrule
\end{tabular}%
}

\vspace{1ex}
\parbox{\textwidth}{
\footnotesize \textbf{PCA:} Principal Component Analysis; \textbf{KDE:} Kernel Density Estimation; \textbf{ML:} Machine Learning; \textbf{RBF:} Radial Basis Function; \textbf{RF:} Random Forest; \textbf{GBR:} Gradient Boosting Regressor; \textbf{Samp.:} Sampling. \\
\textbf{Unconstrained:} Error bands technically cover the true spectrum but are uninformatively wide and violate physical boundaries.\\
\textbf{Overconfident:} Error bands are tight but fail to capture the true spectrum due to latent truncation or missing copulas. \\
\textbf{Calibrated:} Error bands accurately capture statistical variance, though unbounded mathematical frameworks (e.g., Gaussian Copulas) still permit minor physical boundary violations ($y<0$).
}
\end{table*}

\begin{table*}[htbp]
\centering
\caption{Experiment 2: The Hierarchy of Algorithmic Reasoning and Scientific Validity ($\mathcal{V}$)}
\label{tab:reason}
\begin{tabular}{l l l l}
\toprule
\textbf{State} & \textbf{Algorithm /} & \textbf{Reasoning} & \textbf{Scientific Verdict ($\mathcal{V}$)} \\
\textbf{(Input Level)} & \textbf{Architecture} & \textbf{Paradigm} & \\ \midrule

\textbf{T} & Naive / Unconstrained & \textbf{Semantic} & \textbf{$\mathcal{V} = 0$ (Invalid / Fail):} \\
\textit{(The Keyword)} & Generation & \textbf{Guessing} & Fails to execute mathematically or produces \\
 & & & non-physical results that violate basic\\
 & & & domain constraints. \\ \addlinespace \midrule \addlinespace

\textbf{TA} & Bayesian Linear & \textbf{Purely} & \textbf{$\mathcal{V} = 1$ (Unconstrained / Partial):} \\
\textit{(The Idea)} & Regression & \textbf{Mathematical} & Interpolates sparse photometric data \\
 & & & but ignores the physical manifold, leading \\
 & & & to unconstrained boundary violations \\
 & & & and severe overfitting. \\ \addlinespace \midrule \addlinespace

\textbf{TAM} & Independent ML & \textbf{Heuristic /} & \textbf{$\mathcal{V} = 2$ (Overconfident / Functional):} \\
\textit{(The Method)} & (RF / GBR) & \textbf{Generic} & Constrains the mean prediction to the physical \\
 & & & manifold but ignores latent space correlation, \\
 & & & yielding mathematically invalid, independent \\
 & & & sampling geometries. \\ \addlinespace \midrule \addlinespace

\textbf{GT} & AutoGluon + & \textbf{Domain-Specific} & \textbf{$\mathcal{V} = 3$ (Calibrated / Scientific):} \\
\textit{(The Expertise)} & Gaussian Copula & \textbf{(Tacit)} & Respects both the non-linear physical manifold \\
 & & & and the latent covariance, producing physically \\
 & & & constrained, survey-ready statistical coverage. \\ \bottomrule
\end{tabular}

\vspace{1ex}
\parbox{\textwidth}{
\footnotesize \textit{Note.}---This hierarchy illustrates how increasing prompt specificity successfully shifts the LLM through the validity spectrum ($\mathcal{V}=0 \rightarrow 3$). It highlights that while explicit methodological text ($X_{TAM}$) enables functional machine learning ($\mathcal{V}=2$), bridging the final gap to true scientific validity ($\mathcal{V}=3$) requires tacit, domain-specific statistical interventions (e.g., latent copulas).
}
\end{table*}

\subsubsection{Summary of Functional Reproducibility}
\label{sec:exp2_summary}

Synthesizing the qualitative evaluation of frontier API models through our scientific validity hierarchy reveals three key properties of LLM-assisted reproduction:

\begin{enumerate}
    \item \textbf{Conceptual Convergence and the Entropy Floor:} Full methodologies ($X_{TAM}$) reliably constrain models to the correct conceptual framework, achieving functional viability ($\mathcal{V}=2$), but do not enforce convergence to a single implementation. Diverse architectural choices (e.g., Random Forest vs.\ Gradient Boosting) and sampling strategies show that reproducible solutions form a \textit{distributed manifold}. This directly validates the \textbf{Entropy Floor}: functional viability persists even when structural mutual information plateaus.

    \item \textbf{Gap Between Syntactic and Scientific Competence:} Models successfully translate explicit methodology into executable pipelines ($\mathcal{V}=2$), yet consistently fail to enforce implicit physical constraints (e.g., non-negative reflectance, latent covariance). This gap defines the barrier to true scientific calibration ($\mathcal{V}=3$), demonstrating that natural language alone cannot reliably transmit \textbf{implicit expert knowledge}. Functional execution does not guarantee scientific validity.

    \item \textbf{$\mathcal{V}$-Metric Refinement Trajectory:} Reconstruction follows a staged progression along the validity hierarchy. The Title stage ($X_T$) produces generic priors that violate domain constraints ($\mathcal{V}=0$). The Abstract ($X_{TA}$) induces the main conceptual shift but leads to overfitting ($\mathcal{V}=1$). Full text ($X_{TAM}$) adds procedural constraints, reaching functional software ($\mathcal{V}=2$). This trajectory mirrors Experiment 1: late-stage text primarily shifts probability mass (JSD) without reducing uncertainty (near-zero marginal Mutual Information).
\end{enumerate}

These results define a key boundary in computational science: while manuscripts reliably transmit the core algorithmic idea, achieving fully calibrated, physically constrained implementations ($\mathcal{V}=3$) still depends on tacit, domain-specific knowledge rarely captured in text.

\section{Discussion}
\label{sec:disc}

\subsection{Algorithmic Macrostates and the Degeneracy of Implementation}
\label{sec:discussion_macrostate}

We formalize the dynamics of LLM-assisted code generation by mapping our representational spaces onto the statistical mechanics concepts of \textbf{macrostates} and \textbf{microstates}. The \textbf{algorithmic macrostate} represents the overarching conceptual framework prescribed by the manuscript (e.g., combining PCA and KDE to achieve functional viability, $\mathcal{V} \ge 2$). It is the theoretical structure successfully tracked by the lexical space via substantial Mutual Information (MI) gain. The \textbf{algorithmic microstate} is the specific software realization, such as variable assignments, library dependencies, and logic flow, captured in the semantic space. Because the algorithmic hypothesis space is highly degenerate, a single valid macrostate encompasses a vast ensemble of functional microstates. Our results suggest that while natural language effectively extracts MI to constrain the macrostate, it is insufficient to uniquely determine the microstate.

This abstraction helps interpret a persistent challenge in open science: the fragility of direct code reuse. Shared source code represents a single, highly personalized microstate, often entangled with local environments, implicit assumptions, and subjective design choices. As a result, reuse requires reverse-engineering the underlying macrostate from this microstate ``noise.'' Failures of reuse thus often reflect incompatibility of implementations rather than flaws in the scientific method.

This perspective reframes our definition of Ground Truth (Section~\ref{sec:GT}). Reproducibility should not require recovering an identical microstate, but converging to a shared \textbf{algorithmic macrostate}. This allows both human and AI agents to construct functionally equivalent yet structurally distinct implementations.

\subsection{Illusion of Syntactic Competence}
\label{sec:discussion_illusion}

LLMs provide a useful stress test for computational reproducibility. Across both experiments, we observe a practical limitation of natural language in transmitting complete algorithmic structure.

A well-written Methods section is traditionally considered sufficient if a human can reconstruct a functionally equivalent pipeline. However, in the $X_{TAM}$ setting, frontier models that correctly identify the \textit{PCA + KDE + ML} macrostate typically achieve only functional viability ($\mathcal{V}=2$). The gap to full calibration ($\mathcal{V}=3$) arises from missing \textbf{implicit expert knowledge}. For example, while the text specifies density estimation, the physical intuition to prefer non-parametric KDE over GMMs for small TNO samples remains tacit. Likewise, the omission of latent covariance modeling (e.g., via a Gaussian Copula) leads models to adopt uncoupled priors despite correctly assembling the explicit components. This observation is consistent with the view that scientific understanding relies partly on contextual and tacit knowledge beyond explicit text \citep{Ting2026}.

These structural gaps are partially obscured by a persistent \textit{Entropy Floor}: LLMs generate a manifold of valid implementations, producing an algorithmic \textbf{noise floor} (Table~\ref{tab:dis_to_GT}). Despite strong lexical mutual information gain, semantic distance to the Ground Truth remains high (e.g., $0.32$ for \texttt{GPT-5}, $0.34$ for \texttt{Claude}), indicating divergence at the microstate level.

To probe this effect, we performed a targeted ablation by generating $N=200$ implementations with an explicit Gaussian Copula constraint. Qualitative inspection revealed heterogeneous responses: some models incorporated the constraint, while others largely ignored it. However, global metrics (semantic entropy and centroid distance) remained effectively unchanged, and the ensemble persisted as a degenerate cloud in embedding space.

Rather than demonstrating that explicit text cannot reduce structural variance, this result highlights a limitation of our measurement framework. While our semantic embeddings and information-theoretic metrics capture macroscopic conceptual shifts (e.g., $T \rightarrow TA$), they lack sensitivity to localized algorithmic refinements (e.g., $X_{TAM}$ versus $X_{TAM} + \text{Copula}$). Because embeddings aggregate overall structure, they may average out fine-grained statistical constraints. This suggests that the observed entropy floor may arise from two distinct sources: (1) intrinsic degeneracy in the algorithmic hypothesis space, and (2) finite resolution in our semantic measurement framework.

Within the macrostate–microstate framework, this implies that textual constraints can reliably shift the macrostate, while resolving microstate-level variance may require higher-resolution representations beyond natural language and coarse semantic embeddings.

This leaves an open question: whether the observed entropy floor reflects a fundamental limit of text-driven reconstruction or an artifact of the representation and metrics used here. Developing higher-resolution approaches, such as Abstract Syntax Tree divergence or execution-trace alignment, will be essential for resolving this ambiguity.

\begin{table}[htbp]
    \centering
    \small 
    \setlength{\tabcolsep}{6pt} 
    \caption{Distance Metrics to Ground Truth in Lexical and Semantic Spaces}
    \label{tab:dis_to_GT}
    \begin{tabular}{lcc}
        \toprule
        \textbf{Model} & \textbf{Lexical Space} & \textbf{Semantic Space} \\
        \midrule
        GPT-oss  & 0.26 & 0.24 \\
        DeepSeek & 0.16 & 0.18 \\
        Gemini   & 0.14 & 0.13 \\
        GPT-5    & 0.20 & 0.32 \\
        Claude   & 0.20 & 0.34 \\
        \bottomrule
    \end{tabular}
\end{table}

\subsection{The Limits of Textual Verbosity}
\label{sec:discussion_verbosity}

A common response to structural gaps is to increase methodological detail, assuming that more text increases MI and enforces convergence. Our case study suggest this may be inefficient for algorithmic reproducibility.

In Experiment 1, the transition from $X_{TA}$ to $X_{TAM}$ increases lexical constraints but yields near-zero marginal MI in the semantic space ($I(X_{TAM}; Y | X_{TA}) \approx 0$). In some models (e.g., \texttt{GPT-oss}), semantic entropy remains flat, indicating that additional text does not compress the microstate ensemble.

Moreover, increased detail may expand the solution space. Moving from simpler models (e.g., BLR) to highly flexible ML frameworks introduces many valid implementations. If this added complexity does not improve $\mathcal{V}$, it functions as variance rather than constraint. Under \textbf{Occam's razor}, such verbosity constitutes algorithmic over-engineering. Natural language does not behave as a strict compiler; redundant detail reorganizes noise rather than isolating signal.

Taking this application of \textbf{Occam's razor} to its logical conclusion, if additional text fails to meaningfully shift the model's information-theoretic coordinates ($H$, JSD) or improve its scientific validity ($\mathcal{V}$), it should be pruned. In the extreme scenario where the reconstruction from the full manuscript ($X_{TAM}$) is indistinguishable from the title alone ($X_T$), the principle suggests that the entire paper may be redundant. This quantitative plateau perfectly captures \citet{Peiris2026}'s critique of ``scientifically thin'' research, where mechanical execution simply substitutes for thought. 

This motivates a clear criterion for methodological relevance, consistent with the information-theoretic framework of \citet{Fanelli2019}. Information is \textbf{redundant} if frontier LLMs can reliably infer it from standard domain priors or general logic (e.g., routine hyperparameter tuning or standard library imports). As demonstrated by the flat entropy curves in our experiments, documenting these inferable steps adds zero marginal MI. It merely clutters the narrative with mechanical execution, diluting the scientific signal.

Conversely, information is \textbf{essential} when it actively forces the algorithm to deviate from generic statistical priors to respect the actual physics of the domain. In our case study, providing explicit mathematical components (like PCA or ML) was insufficient to achieve scientific calibration ($\mathcal{V}=3$). The true essential links were the \textbf{structural constraints} driven by tacit expert judgment: the specific physical intuition to select non-parametric KDE over parameterized GMMs for highly limited TNO samples, and the necessity of a Gaussian Copula to preserve latent physical covariance. 

These non-inferable constraints represent the irreducible information: the ``integrated physical intuition'' \citep{Peiris2026} that successfully guides the algorithmic microstate toward the solution. This directly underpins the \textit{Primacy of the Idea} (Section~\ref{sec:exp1_summary}): the initial conceptual leap ($T \rightarrow TA$) typically generates the largest Information Gain and JSD shift because it is the stage where this crucial, domain-specific intuition is first introduced. To maximize reproducibility and scientific value, authors should explicitly center these essential, symmetry-breaking decisions, while ruthlessly pruning the inferable mechanical noise. This raises a practical question: which reproducibility artifacts can efficiently transmit these non-inferable constraints without simply increasing textual verbosity?

\subsection{Pseudocode and Layered Reproducibility Artifacts}
A natural mitigation strategy is to provide pseudocode in method papers.
This approach is efficient because it constrains the algorithmic macrostate: it fixes the ordering of operations, the inputs and outputs, and the major modeling components. It should therefore reduce lexical ambiguity and eliminate broad families of incorrect reconstructions. However, pseudocode alone does not necessarily determine the implementation microstate. If it merely reformats explicit prose into procedural syntax, while leaving physical priors, calibration choices, numerical conventions, software defaults, or symmetry-breaking decisions implicit, the semantic entropy floor remains. In that case, an LLM can generate executable and superficially plausible code that still fails to reach scientific calibration. We therefore view pseudocode as a valuable but incomplete artifact. Its reproducibility value is maximized when paired with explicit domain constraints, minimal executable reference implementations, and benchmark intermediate outputs that constrain not only the intended algorithm, but also its acceptable numerical and physical behavior.

\subsection{AI as a Diagnostic Tool for Open Science}
\label{sec:discussion_diagnostic}

The tendency of LLMs to ``brain-fill'' missing procedural details can be repurposed as a diagnostic test for methodological clarity. If an LLM can reliably infer a missing step from the manuscript and standard domain priors, then that step likely belongs to the inferable background knowledge of the field. Conversely, if multiple frontier LLMs fail to infer the same step, or consistently replace it with a generic but scientifically incorrect prior, this failure identifies a non-inferable constraint that must be explicitly documented. We propose using LLMs as a \textbf{Zero-Shot Stress Test} for methodological clarity.

By performing a \textbf{blind algorithmic reproduction} prior to submission, authors can task a frontier LLM to reconstruct the implementation strictly from the manuscript text. In practice, this stress test can be implemented as a simple protocol: (1) generate an ensemble of independent reconstructions from the manuscript, (2) evaluate each reconstruction under a functional or domain-specific validity criterion (e.g., $\mathcal{V}$), and (3) identify consistent failure modes across the ensemble. These recurring failures mark the locations where the model fails to reach scientific calibration ($\mathcal{V} = 3$), providing a high-resolution diagnostic of hidden assumptions and missing structural constraints.

These failure points flag the implicit expert knowledge (e.g., the necessity of a Copula) that human reviewers, often guided by shared domain intuition, may overlook. Conversely, this stress test identifies redundant textual descriptions that yield zero marginal Mutual Information or inter-state shift, reinforcing the principle of algorithmic Occam's razor.

We further propose the \textbf{reproducibility prompt set} as a complementary artifact. Recent editorial perspectives have emphasized the need for practical, community-wide standards to safeguard the integrity of astronomical research \citep{NatAstronEd2026}, including the use of prompt and response transcripts in related disciplines. Our proposal directly aligns with this direction. While source code captures a single microstate, the prompt set encodes the macrostate: structural constraints, priors, and essential assumptions. Together, they provide a dual representation: a historical implementation and a generative blueprint that supports reproducibility across evolving computational environments.

Although derived from a single case study, this pattern likely extends to broader astrophysical and computational workflows, where statistical frameworks are coupled with partially documented domain constraints. As LLMs become increasingly embedded in research pipelines, reliance on tacit knowledge becomes a growing vulnerability. Formalizing these constraints through prompt-based artifacts offers a practical pathway to preserve scientific integrity in an increasingly automated research ecosystem.

\section{Summary}
\label{sec:summary}

As artificial intelligence increasingly intersects with scientific research, Large Language Models offer a compelling new mechanism for computational reproducibility. In this study, we introduced a novel information-theoretic framework to quantify methodological transparency, measuring the effective volume of algorithmic information successfully transmitted by a scientific manuscript. We applied this framework to an astronomical pipeline for Trans-Neptunian Object (TNO) spectral reconstruction, utilizing frontier LLMs to reconstruct the codebase using only the methodological text provided. Through our dual-experiment design, we measured both the theoretical reduction of uncertainty in the algorithmic hypothesis space and the practical scientific validity of the generated software.

Our case study reveal that while explicit methodological text successfully constrains the overarching algorithmic approach, the uncertainty of reproduction does not compress to zero. By mapping solutions into lexical and semantic embedding spaces, we demonstrated that reproducibility operates on two distinct levels: the conceptual \textbf{macrostate} and the structural \textbf{microstate}. While textual information successfully collapses macroscopic uncertainty, the specific implementation remains fundamentally \textit{degenerate}. Functionally reproducible code exists as a distributed manifold of varied \textit{algorithmic dialects}, architectural choices, and library dependencies, empirically supporting the existence of an apparent \textbf{Entropy Floor} within our experimental scope.

Our physical code generation experiment highlights a critical boundary in computational astrophysical methodology: the \textbf{illusion of syntactic competence}. While LLMs excel at filling in \textit{hidden implementation details} to produce executable code, they systematically struggled to bridge the \textbf{implicit knowledge gap}. Without explicit instructions, models failed to enforce the unwritten physical constraints of the domain, such as bounding positive reflectances and preserving the empirical data manifold via Gaussian Copulas. Consequently, models produced pipelines that were mathematically functional but scientifically invalid.

While derived from a single astrophysical case study, these results point toward a necessary evolution in how we approach Open Science. The traditional gold standard of sharing raw source code merely transmits a single, often brittle \textit{microstate}. Conversely, a rigorously detailed manuscript transmits a robust conceptual \textit{macrostate}. If authors explicitly formalize their implicit domain knowledge within the text, and supplement it with reproducibility prompt sets, LLMs have the potential to act as versatile compilers, translating that shared macrostate into a native, functional microstate adapted to the reproducing researcher's computational environment. 

To achieve this, we encourage the scientific community to explore LLMs as proactive auditors of scientific transparency. By subjecting drafts to a ``Zero-Shot Stress Test'': prompting models to execute blind algorithmic reproductions, researchers can systematically expose hidden assumptions and unstated physical boundaries. This approach offers a pathway to transform subjective physical intuition into a more universally reproducible methodology, moving the field toward a transparent and resilient open science.

\section*{Software and Data Availability}

The source code and the complete Reproducibility Prompt Set for the information-theoretic framework and Experiment 1 are publicly available on GitHub at \url{https://github.com/sevenlin123/astro_entropy}. 

The generative spectral reconstruction pipeline applied in our case study, including the trained PCA manifold, is fully open-source. The source code and demonstration notebooks can be accessed on GitHub at \url{https://github.com/sevenlin123/spectra_reconstructor/}.

\textit{Software used in data analysis.} The LLM-based reconstruction experiments used the following models: \texttt{DeepSeek-R1-Distill-Qwen-14B} \citep[\texttt{DeepSeek},][]{deepseek2025r1}, \texttt{GPT-oss-20b} \citep[\texttt{GPT-oss},][]{openai2025gptoss}, \texttt{Gemini 2.5 Pro} \citep[\texttt{Gemini},][]{google2025gemini25pro}, \texttt{Gemini 3.1 Pro} \citep[\texttt{Gemini},][]{google2026gemini31pro}, \texttt{ChatGPT 5.2} \citep[\texttt{GPT},][]{openai2026gpt52}, and \texttt{Claude 4.6 Sonnet} \citep[\texttt{Claude},][]{anthropic2026claude46}. Where no DOI is available, the cited model card or official provider documentation gives the access point used to identify the model.

\begin{acknowledgments}

 We thank Ming-Liang Lin for valuable discussions on Polanyi's concept of tacit knowledge. We also thank the anonymous referee for constructive suggestions that improved the clarity and presentation of this manuscript.
 This material is based upon work supported by the National Science Foundation under grant No. AST-2406527.

\end{acknowledgments}

\appendix

\section{Methodological Keyword Whitelist}
\label{app:keywords}

To quantify lexical convergence and calculate the information-theoretic metrics (Entropy and JSD) presented in Experiment 1, we defined a controlled vocabulary of 57 domain-specific keywords. This whitelist was designed to capture the core methodological components of the TNO spectral reconstruction pipeline while excluding mundane English stop words or non-scientific syntax. The keywords are categorized below:

\begin{itemize}
    \item \textbf{Machine Learning \& Clustering:} \textit{dbscan, kmeans, random forest, svm, support vector, neural network, xgboost, gradient boosting, gbm, gbdt, unsupervised, supervised, manifold, generative, data augmentation, autoencoders, knn, gan, diffusion, t-sne.}
    \item \textbf{Statistical Modeling \& Regression:} \textit{least squares, curve fit, chi square, levenberg, marquardt, linear regression, regression, linear, logistic regression, logistic, copula, linalg, influence.}
    \item \textbf{Probabilistic Sampling \& Inference:} \textit{mcmc, markov chain, monte carlo, gp, gaussian process, bayesian, emcee, likelihood, forward modeling, inverse problem, variational inference, sampling, importance sampling, rejection sampling, metropolis, hastings.}
    \item \textbf{Dimensionality Reduction \& Analysis:} \textit{principal component analysis, pca, kernel density estimation, kde, components, component, eigenvalues, eigenvectors.}
\end{itemize}

This restricted vocabulary defines the basis of the probability distributions used to measure the information-theoretic distance between the generated solutions and the Ground Truth.

\section{Ground Truth Reference Implementation}
\label{app:ground_truth}

To establish the degenerate ground-truth distribution $P_{\mathrm{GT}}(Y)$ defined in Section~\ref{sec:GT} and implement in Section~\ref{sec:metric}, the original authors' validated computational pipeline was distilled into a structured algorithmic and pseudocode format. This extraction was performed using \texttt{Gemini 2.5 Pro} directly on the case study's source code repository, followed by strict manual verification by the original author to ensure zero hallucination and perfect methodological fidelity.

The resulting Ground Truth algorithm, which serves as the quantitative anchor for all Jensen-Shannon Divergence and Cosine Distance alignment metrics in Experiment 1 and Experiment 2, is provided below.

\begin{tcolorbox}[colback=blue!5!white, colframe=blue!75!black, title=Algorithm: Ground Truth Spectral Reconstruction, arc=3mm, breakable]

\textbf{Phase 1: Manifold Learning and Data Augmentation}
\begin{enumerate}
    \item \textbf{Spectral Compression (PCA):} High-dimensional spectral data ($D \approx 850$ points) is projected onto a lower-dimensional ``latent space'' using Principal Component Analysis. Typically, 9 Principal Components (PCs) are used, capturing over 97\% of the spectral variance.
    \item \textbf{Density Estimation (KDE):} The distribution of real spectra in this latent space is modeled using Kernel Density Estimation (KDE). This creates a ``manifold'' of physically plausible spectral shapes.
    \item \textbf{Synthetic Data Generation:} To train robust models, thousands of synthetic spectra are generated by sampling from the KDE manifold and applying the inverse PCA transform.
\end{enumerate}

\textbf{Phase 2: Photometry Simulation}
\begin{enumerate}
    \item \textbf{Filter Convolution:} Synthetic photometry is generated by convolving spectra with NIRCam throughput curves (managed by \texttt{NIRCamFilters}).
    \item \textbf{Observation Noise:} Realistic observational uncertainty is added to the synthetic magnitudes (e.g., Gaussian noise) to simulate telescope performance.
\end{enumerate}

\textbf{Phase 3: Reconstruction Model Training}
\begin{enumerate}
    \item \textbf{Multi-Target Regression:} An AutoGluon ensemble (using Gradient Boosting Machines) is trained to map from Photometric Magnitudes (Inputs) $\rightarrow$ PCA Components (Targets).
    \item \textbf{Quantile Regression for Uncertainty:} Instead of simple point estimates, the models perform quantile regression (predicting 99 quantiles from 0.01 to 0.99). This allows the system to estimate the full probability distribution of the spectral shape.
\end{enumerate}

\textbf{Phase 4: Probabilistic Reconstruction (Monte Carlo)}
\begin{enumerate}
    \item \textbf{Residual Correlation:} Since errors in predicting different PCs are often correlated, a global correlation matrix (calculated from validation residuals) is used to maintain physical consistency.
    \item \textbf{Correlated Sampling:}
    \begin{itemize}
        \item Draw samples from a multivariate normal distribution using the correlation matrix.
        \item Transform these into uniform probabilities via the Normal CDF.
        \item Map these probabilities to physical PC values using the Inverse CDF (interpolating across the predicted AutoGluon quantiles).
    \end{itemize}
    \item \textbf{Inverse PCA Transform:} The sampled PC vectors are transformed back into the original wavelength space to produce a family of possible spectra.
\end{enumerate}

\textbf{Summary of Workflow}
\begin{enumerate}
    \item \textbf{Input:} Photometric magnitudes from NIRCam filters.
    \item \textbf{Step 1:} Predict 99 quantiles for each of the 9 PCA components.
    \item \textbf{Step 2:} Generate 1,000 correlated samples in the PC latent space.
    \item \textbf{Step 3:} Perform Inverse PCA to get 1,000 spectral realizations.
    \item \textbf{Output:} Reconstructed mean spectrum, and 95\% confidence interval.
\end{enumerate}

\vspace{0.5cm}
\hrule
\vspace{0.5cm}

\begin{verbatim}
// =====================================================================
// Phase 1: Manifold Learning and Data Augmentation
// =====================================================================
FUNCTION generate_synthetic_manifold(real_spectra, num_pcs=9, num_samples=10000):
    // 1. Spectral Compression
    // Project high-dimensional spectra (e.g., 850 points) to latent space
    pca_model = train_PCA(real_spectra, n_components=num_pcs)
    real_latent_pcs = pca_model.transform(real_spectra)
    
    // 2. Density Estimation
    // Model the physical distribution of real spectra in the latent space
    kde_manifold = train_KDE(real_latent_pcs)
    
    // 3. Synthetic Data Generation
    // Sample physically plausible intermediate states
    synthetic_latent_pcs = kde_manifold.sample(num_samples)
    synthetic_spectra = pca_model.inverse_transform(synthetic_latent_pcs)
    
    RETURN pca_model, synthetic_latent_pcs, synthetic_spectra

// =====================================================================
// Phase 2: Photometry Simulation
// =====================================================================
FUNCTION simulate_observations(synthetic_spectra, nircam_filters, noise_model):
    // 1. Filter Convolution
    // Integrate spectra over specific NIRCam throughput curves
    synthetic_mags = convolve_spectra_with_filters(synthetic_spectra, nircam_filters)
    
    // 2. Observation Noise
    // Add realistic telescope uncertainty to generate training inputs
    noisy_mags = apply_gaussian_noise(synthetic_mags, noise_model)
    
    RETURN noisy_mags

// =====================================================================
// Phase 3: Reconstruction Model Training
// =====================================================================
FUNCTION train_reconstruction_pipeline(noisy_mags, synthetic_latent_pcs, validation_data):
    // 1 & 2. Multi-Target Quantile Regression
    // Map Photometry -> 99 Quantiles for each PCA component
    target_quantiles = generate_array(0.01 to 0.99, step=0.01)
    
    autogluon_ensemble = train_GBM_ensemble(
        inputs = noisy_mags,
        targets = synthetic_latent_pcs,
        quantiles = target_quantiles
    )
    
    // Compute residual correlation from validation set for physical consistency
    predictions_val = autogluon_ensemble.predict(validation_data.mags)
    residuals = validation_data.pcs - extract_median(predictions_val)
    global_correlation_matrix = compute_correlation(residuals)
    
    RETURN autogluon_ensemble, global_correlation_matrix

// =====================================================================
// Phase 4 & Summary: Probabilistic Reconstruction (Monte Carlo)
// =====================================================================
FUNCTION reconstruct_spectrum(obs_mags, model_pipeline, pca_model, num_samples=1000):
    autogluon_ensemble = model_pipeline.ensemble
    corr_matrix = model_pipeline.correlation_matrix
    
    // Step 1: Predict 99 quantiles for each of the 9 PCA components
    predicted_quantiles = autogluon_ensemble.predict_quantiles(obs_mags)
    
    // Step 2: Generate correlated samples in the PC latent space
    // 2a. Draw from multivariate normal using pre-calculated correlation matrix
    mv_normal_samples = draw_multivariate_normal(mean=0, cov=corr_matrix, size=num_samples)
    
    // 2b. Transform to uniform probabilities
    uniform_probs = normal_CDF(mv_normal_samples)
    
    // 2c. Map to physical PC values via Inverse CDF interpolation
    sampled_pcs = interpolate_inverse_CDF(uniform_probs, predicted_quantiles)
    
    // Step 3: Perform Inverse PCA
    // Map the 1,000 latent space samples back to the high-dimensional space
    reconstructed_spectra_family = pca_model.inverse_transform(sampled_pcs)
    
    // Output calculation
    reconstructed_mean = calculate_mean(reconstructed_spectra_family, axis=0)
    confidence_interval_95 = calculate_percentiles(reconstructed_spectra_family,
                          [2.5, 97.5], axis=0)
    
    RETURN reconstructed_mean, confidence_interval_95
\end{verbatim}
\end{tcolorbox}

\bibliography{PASPsample701}{}
\bibliographystyle{aasjournalv7}

\end{document}